\documentclass[prd,twocolumn,nofootinbib,superscriptaddress]{revtex4-2}
\usepackage{lipsum}
\usepackage{amsmath,amssymb,mathtools}
\usepackage{mathrsfs}
\usepackage{tensor}
\usepackage{color,xcolor}
\usepackage{hyperref}
\hypersetup{colorlinks=true,
            linkcolor=magenta,
            filecolor=black,
            urlcolor=magenta,
            citecolor=magenta
           }
\allowdisplaybreaks
\makeatletter

\newcommand{\Rmnum}[1]{\expandafter\@slowromancap\romannumeral #1@}
\makeatother
\def\Hunnu{Department of Physics, Institute of Interdisciplinary Studies, Hunan Research Center of the Basic Discipline for Quantum Effects and Quantum Technologies, Key Laboratory of Low Dimensional Quantum Structures and Quantum Control of Ministry of Education, Synergetic Innovation Center for Quantum Effects and Applications, Hunan Normal University,  Changsha, Hunan 410081, People's Republic of China}
\def\Yangzhou{Center for Gravitation and Cosmology, College of Physical Science and Technology, Yangzhou University, Yangzhou 225009, People's Republic of China}
\begin{document}
\title{Vacuum breakdown in a misaligned magnetized Kerr spacetime}
\author{Ruixin Yang}%\email{202410110239@hunnu.edu.cn}
\affiliation{\Hunnu}
\author{Songbai Chen}\email{csb3752@hunnu.edu.cn \textcolor{black}{(corresponding author)}}
\affiliation{\Hunnu}\affiliation{\Yangzhou}
\author{Jiliang Jing}\email{jljing@hunnu.edu.cn}
\affiliation{\Hunnu}\affiliation{\Yangzhou}
%\date{\today}
%%%%%%%%%%%%%%%%%%%%%%%%%%%%%%%%%%%%%%%%%%%%%%%%%%%%%%%%%%%%%%%%%%%%%%%%%%%%
\begin{abstract}
Electron-positron ($e^{+}e^{-}$) pair creation by vacuum breakdown around compact objects is believed to power high-energy astrophysical transients like gamma-ray bursts (GRBs). In this work, we focus on vacuum breakdown around a Kerr black hole (BH) immersed in an asymptotically uniform magnetic field that is inclined with respect to the BH spin axis. The dyadoregion, the region where the induced electric field exceeds the critical value $E_{\text{c}}=m_{e}^{2}c^{3}/(e\hbar)$, is identified via the electromagnetic invariants. It is found that the dyadoregion consists of several lobes whose number, size, and orientation vary with the inclination. We also estimate the electromagnetic energy available for pair creation and derive a beaming factor that allows a conversion between the intrinsic dyadoregion energy and the observed isotropic energy. The thermodynamic properties of the resulting electron-positron-photon ($e^{+}e^{-}\gamma$) plasma are included, revealing an initial magnetic dominance. The evaluation of the minimum magnetic field required shows that misaligned magnetic fields generally favor pair creation more than aligned ones.
\end{abstract}
%%%%%%%%%%%%%%%%%%%%%%%%%%%%%%%%%%%%%%%%%%%%%%%%%%%%%%%%%%%%%%%%%%%%%%%%%%%%
\maketitle
%%%%%%%%%%%%%%%%%%%%%%%%%%%%%%%%%%%%%%%%%%%%%%%%%%%%%%%%%%%%%%%%%%%%%%%%%%%%
\section{Introduction}\label{sec:intro}
In quantum electrodynamics, when an electric field exceeds the critical value $E_{\text{c}}=m_{e}^{2}c^{3}/(e\hbar)\approx 4.41\times 10^{13}\operatorname{G}$, the vacuum becomes unstable and decays into $e^{+}e^{-}$ pairs via Schwinger mechanism \cite{Schwinger:1951nm}. While such an extreme field strength is challenging to achieve in terrestrial laboratories, it may naturally occur in the surroundings of compact astrophysical objects. In particular, around stellar-mass BHs, the interplay of rotation and external magnetic fields can induce electric fields of sufficient magnitude to trigger this vacuum breakdown, leading to the copious production of $e^{+}e^{-}$ pairs \cite{Ruffini:2009hg}. This process offers a possible physical origin for the $e^{+}e^{-}\gamma$ plasma that is believed to power the prompt emission of GRBs. Compared to traditional GRB models, which often rely on neutrino annihilation from a massive accretion disk \cite{Popham:1998ab,Narayan:2001qi,Kohri:2002kz,DiMatteo:2002iex,Kohri:2005tq,Lee:2005se,Gu:2006nu,Chen:2006rra,Kawanaka:2007sb,Janiuk:2009gc,Kawanaka:2012ub,Luo:2013zx,Xue:2013boa,Liu:2017kga} and require extremely high accretion rates \cite{Chen:2006rra,Liu:2017kga,Uribe:2019cpq}, the vacuum breakdown operates on a much shorter timescale \cite{Aksenov:0227,Aksenov:2009dy} and can efficiently convert the electromagnetic energy into the pair plasma. This inherent efficiency positions it as a viable alternative for explaining the central engine of high-energy transients.

For a physically acceptable BH immersed in an electromagnetic field, vacuum breakdown does not happen over the entire space. Therefore, it is necessary to identify the so-called dyadoregion\footnote{The term ``dyado'' comes from the Greek word \textit{duas-duados}, meaning pairs \cite{Preparata:1998rz}.}, where, outside the event horizon, the electric field strength (measured by a specific observer family) exceeds the critical value $E_{\text{c}}$ and $e^{+}e^{-}$ pairs are created. The dyadoregion of Reissner-Nordstr\"{o}m BHs with an electric monopole is just a spherical shell supported by $r_{+}\leqslant r\leqslant r_{\text{d}}$, with $r_{+}=M+\sqrt{M^{2}-Q^{2}}$ and $r_{\text{d}}=\sqrt{\left\lvert Q\right\rvert/E_{\text{c}}}$ \cite{Preparata:1998rz}, where $M$ and $Q$ are the BH mass and charge, respectively. For the case of Kerr-Newman BHs, however, it has been shown in \cite{Cherubini:2009ww} that the boundary of the dyadoregion exhibits a toroidal topology for certain parameters.

Actually, even when a BH carries no net charge, the vacuum in its vicinity can also be polarized. The most typical example is a Kerr BH immersed in an asymptotically uniform test magnetic field aligned with the BH spin axis at infinity, as considered by Wald \cite{Wald:1974np}. In this configuration, the dragging of magnetic field lines by the rotating BH induces a quadrupolar electric field, which is responsible for initiating vacuum breakdown. This is relevant because, in realistic astrophysical environments, BHs are generally expected to be electrically neutral.

Within the framework of the Wald solution, Cherubini et al. \cite{Cherubini:2025lnc} performed a systematic study of vacuum breakdown around a stellar-mass Kerr BH. They provided an invariant characterization of the dyadoregion, detailing its morphology, the electromagnetic energy available for pair creation, and the thermodynamic properties of the pair plasma. It is found that the electromagnetic energy is mostly concentrated in the two polar lobes along the BH's rotation axis, with a semi-aperture angle $\approx 30.42^{\circ}$ \cite{Cherubini:2025lnc}. These results constituted the initial conditions for simulations of plasma dynamics in transient astrophysical events like GRBs. In fact, the Wald solution with an overcritical electric field has been used as the inner engine to model the prompt emission of GRBs, see, e.g., Refs.~\cite{Moradi:2021hus,Rastegarnia:2022rds}. In those works, however, the electromagnetic energy stored in the dyadoregion and the pair plasma parameters were estimated by using the effective charge method \cite{Cherubini:2009ww}.

One of the key assumptions in \cite{Cherubini:2025lnc} is that the external magnetic field is parallel to the BH spin axis. However, such a perfect alignment is unlikely to hold in astrophysical reality. The progenitor systems that lead to a stellar-mass BH formation---such as the core collapse of a massive star or the merger of binary neutron stars (NSs)---typically involve complex, turbulent, and differentially rotating magnetized fluids \cite{Obergaulinger:2008pb,Obergaulinger:2017qno,Kiuchi:2014hja,Kiuchi:2015sga}. In these processes, the magnetic field inherited by the BH may be tilted relative to its rotation axis due to asymmetric accretion, off-axis jet launching, or the misalignment between the angular momentum of the collapsing core and the large-scale ambient field \cite{Proga:2003ap,Liska:2017alm}. Moreover, wind-fed BHs, NS-BH binaries, high-mass X-ray binaries, Gaia BHs \cite{El-Badry:2023pah}, and isolated BHs accreting from the interstellar medium \cite{Barkov:2012sj,Kin:2025axi} are also expected to host inclined magnetic fields \cite{Figueiredo:2025xbo}. Fundamentally, since the accretion flow falling toward a BH from large distances is unaware of the BH's spin direction, a net inclination may persist farther away even if the magnetic field becomes aligned with the BH spin near the horizon via the magneto-spin alignment mechanism \cite{McKinney:2012wd}. This motivates us to go beyond the aligned case and investigate vacuum breakdown around a rotating BH immersed in an arbitrarily oriented test magnetic field.

In Kerr background, the solution to Maxwell's equations generated by a magnetic field which is uniform at spatial infinity but not aligned with the BH rotation axis was first obtained by Bi{\v{c}}{\'a}k and Dvo{\v{r}}{\'a}k \cite{Bicak1976} using Newman-Penrose formalism. The nonvanishing components of the corresponding electromagnetic field tensor in Boyer-Lindquist coordinates were explicitly given by Bičák and Janiš \cite{Bicak1985}. The inclined magnetic field configuration has been extensively studied so far, including the discovery of magnetic null points and reconnection sites \cite{Karas:2008xj,Karas:2012mp}, the induction of chaos in charged particle dynamics \cite{Kopacek:2014moa,Kopacek:2014ooa}, the monarch-butterfly-like migration of Carrollian particles on the magnetized horizon \cite{Bicak:2023rsz}, and the latest 3D kinetic simulations revealing the effect of inclination on jet power and particle acceleration \cite{Figueiredo:2025xbo}. For further interesting astrophysical applications of this subject, see Refs.~\cite{Karas:1989zz,Punsly:1989zz,Dovciak_2000,Neronov:2007vy,Neronov:2009zz,Kalashev:2012cm,Tursunov:2019mox,Karas:2020ixz,Kopacek:2020scv,Ressler:2021jjr,Chen:2021sya,Hu:2021paa,James:2024bib,Frolov:2024xyo,Frolov:2024tiu}. To the best of our knowledge, a systematic study of vacuum breakdown around a Kerr BH immersed in a misaligned magnetic field has not yet been carried out.

The remainder of this paper is organized as follows. In the next section, we briefly review the solution for an asymptotically uniform and inclined electromagnetic field in Kerr spacetime to set the stage. In Sec.~\ref{sec:dyadoregion}, we identify the dyadoregion and calculate the electromagnetic energy stored inside. It will be shown that breaking the axial symmetry leads to a more diverse morphology for the dyadoregion. The thermodynamic properties of the pair plasma are provided in Sec.~\ref{sec:thermo}. Finally, Sec.~\ref{sec:conclu} presents a summary of our main results.
%%%%%%%%%%%%%%%%%%%%%%%%%%%%%%%%%%%%%%%%%%%%%%%%%%%%%%%%%%%%%%%%%%%%%%%%%%%%
\medskip
\section{Kerr BH placed in a misaligned magnetic field}\label{sec:testfield}
Magnetic fields play a crucial role in astrophysics. It is important to note that all the magnetic fields observed in the vicinity of astrophysical objects are so weak in the sense that their energy density is negligible compared to the spacetime curvature \cite{Tursunov:2019oiq}. Indeed, the inequality \cite{Galtsov:1978ag}
\begin{equation}
B\ll 2.4\times 10^{19}\left(\frac{M_{\odot}}{M}\right)\operatorname{G}
\end{equation}
is satisfied for any such system. This implies that we can work in a fixed background without taking into account the effect of the magnetic field on the geometry.

The line element of the Kerr metric in Boyer-Lindquist coordinates $\{t,r,\theta,\varphi\}$ is given by
\begin{align}\label{eq:metric}
ds^{2}=&-\frac{\Delta-a^{2}\sin^{2}\theta}{\rho^{2}}dt^{2}+\frac{\rho^{2}}{\Delta}dr^{2}+\rho^{2}d\theta^{2}\notag\\
&+\frac{(r^{2}+a^{2})^{2}-\Delta a^{2}\sin^{2}\theta}{\rho^{2}}\sin^{2}\theta d\varphi^{2}\notag\\
&-2a\frac{r^{2}+a^{2}-\Delta}{\rho^{2}}\sin^{2}\theta dtd\varphi,
\end{align}
with
\begin{align}
\rho^{2}&=r^{2}+a^{2}\cos^{2}\theta,\\
\Delta&=r^{2}-2Mr+a^{2},
\end{align}
where $M$ and $a$ are the BH mass and spin, respectively. As is well known, the solution \eqref{eq:metric} admits two linearly independent Killing vectors
\begin{equation}\label{eq:Killing}
\xi=\frac{\partial}{\partial t},\quad\psi=\frac{\partial}{\partial\varphi}.
\end{equation}
The event horizon is located at
\begin{equation}\label{eq:horizon}
r_{+}=M+\sqrt{M^{2}-a^{2}}.
\end{equation}

When a Kerr BH is placed in a uniform magnetic field, the original field is deformed by the hole, and the resulting electromagnetic field reads \cite{Bicak1976,Bicak1985}
\begin{widetext}
\begin{align}
A_{t}&=-aB_{0}\left[1-\frac{Mr}{\rho^{2}}(1+\cos^{2}\theta)\right]+\frac{aM}{\rho^{2}}\sin\theta\cos\theta\left[B_{1}(r\cos\phi-a\sin\phi)+B_{2}(r\sin\phi+a\cos\phi)\right],\label{eq:A0}\\
A_{r}&=-(r-M)\sin\theta\cos\theta(B_{1}\sin\phi-B_{2}\cos\phi),\label{eq:A1}\\
A_{\theta}&=-a(r\sin^{2}\theta+M\cos^{2}\theta)(B_{1}\cos\phi+B_{2}\sin\phi)-\left[r^{2}\cos^{2}\theta-(Mr-a^{2})\cos2\theta\right](B_{1}\sin\phi-B_{2}\cos\phi),\label{eq:A2}\\
A_{\varphi}&=\frac{1}{2}B_{0}\sin^{2}\theta\left[r^{2}+a^{2}-\frac{2a^{2}Mr}{\rho^{2}}(1+\cos^{2}\theta)\right]\notag\\
&\quad\hspace{1pt}-\sin\theta\cos\theta\left\{\Delta(B_{1}\cos\phi+B_{2}\sin\phi)+\frac{M(r^{2}+a^{2})}{\rho^{2}}\left[B_{1}(r\cos\phi-a\sin\phi)+B_{2}(r\sin\phi+a\cos\phi)\right]\right\},\label{eq:A3}
\end{align}
\end{widetext}
where $B_{0},B_{1},B_{2}$ are constants, and
\begin{equation}\label{eq:phi}
\phi=\varphi+\int\frac{a}{\Delta}dr
\end{equation}
is the azimuthal angle in Kerr ingoing coordinates. The Faraday tensor can be evaluated by
\begin{equation}\label{eq:Faraday}
F_{\mu\nu}=\partial_{\mu}A_{\nu}-\partial_{\nu}A_{\mu}.
\end{equation}
It is straightforward to verify that $F_{\mu\nu}$ satisfies the source-free Maxwell equations
\begin{equation}
\partial_{\mu}\!\left(\sqrt{-g}F^{\mu\nu}\right)=0,
\end{equation}
with $\sqrt{-g}=\rho^{2}\sin\theta$. Moreover, $F_{\mu\nu}$ is stationary but not axisymmetric, i.e.,
\begin{equation}\label{eq:notaxisymmetric}
\mathcal{L}_{\xi}F_{\mu\nu}=0,\quad\mathcal{L}_{\psi}F_{\mu\nu}\neq 0,
\end{equation}
where $\mathcal{L}$ denotes the Lie derivative operator. To clarify the physical meaning of constants $B_{0}$, $B_{1}$, and $B_{2}$ in the vector potential above, one may examine the behavior of the electromagnetic field \eqref{eq:Faraday} as $r\to\infty$. The result in asymptotic Lorentzian coordinates is simply
\begin{equation}
F_{\mu\nu}\approx
\begin{bmatrix}
0&0&0&0\\
0&0&B_{0}&-B_{2}\\
0&-B_{0}&0&B_{1}\\
0&B_{2}&-B_{1}&0\\
\end{bmatrix}
,
\end{equation}
which indicates that $B_{0}$, $B_{1}$, and $B_{2}$ are the components of the original uniform magnetic field along the $z$-, $x$-, and $y$-axes, respectively, where the $z$-axis is aligned with the BH spin axis. If $B_{1}=B_{2}=0$, in the absence of the perpendicular components, the Wald solution \cite{Wald:1974np}
\begin{equation}\label{eq:Wald}
A_{\mu}=B_{0}\left(a\xi_{\mu}+\frac{1}{2}\psi_{\mu}\right)
\end{equation}
is recovered from Eqs.~\eqref{eq:A0}--\eqref{eq:A3}, as expected.

Hereafter, without loss of generality, we shall always set $B_{2}=0$. For convenience, we also define the inclination angle $i$ between the magnetic field and the BH spin axis as
\begin{equation}
B_{0}=B\cos i,\quad B_{1}=B\sin i,
\end{equation}
with the asymptotic field strength $B=\sqrt{B_{0}^{2}+B_{1}^{2}}$.

Thus far, we have only focused on the geometric objects in Kerr spacetime. It is now appropriate to introduce a family of preferred observers, the zero angular momentum observers (ZAMOs) with four-velocity
\begin{equation}\label{eq:ZAMO}
Z=\sqrt{-g^{tt}}\left(\xi+\omega\psi\right),\quad\omega=-\frac{g_{t\varphi}}{g_{\varphi\varphi}},
\end{equation}
which is orthogonal to the hypersurfaces of constant Killing time $t$. The electric and magnetic fields measured by a ZAMO are, respectively, defined as follows
\begin{equation}\label{eq:ZAMOEandB}
E_{\mu}=F_{\mu\nu}Z^{\nu},\quad B_{\mu}=-{}^{\ast\!}F_{\mu\nu}Z^{\nu},
\end{equation}
where the dual 2-form
\begin{equation}
{}^{\ast\!}F_{\mu\nu}=\frac{1}{2}\varepsilon_{\mu\nu\sigma\rho}F^{\sigma\rho},
\end{equation}
with $\varepsilon_{\mu\nu\sigma\rho}$ being the Levi-Civita tensor.

\begin{figure*}[htbp]
\centering
\begin{minipage}[b]{\textwidth}
\centering
\includegraphics[width=0.32\textwidth]{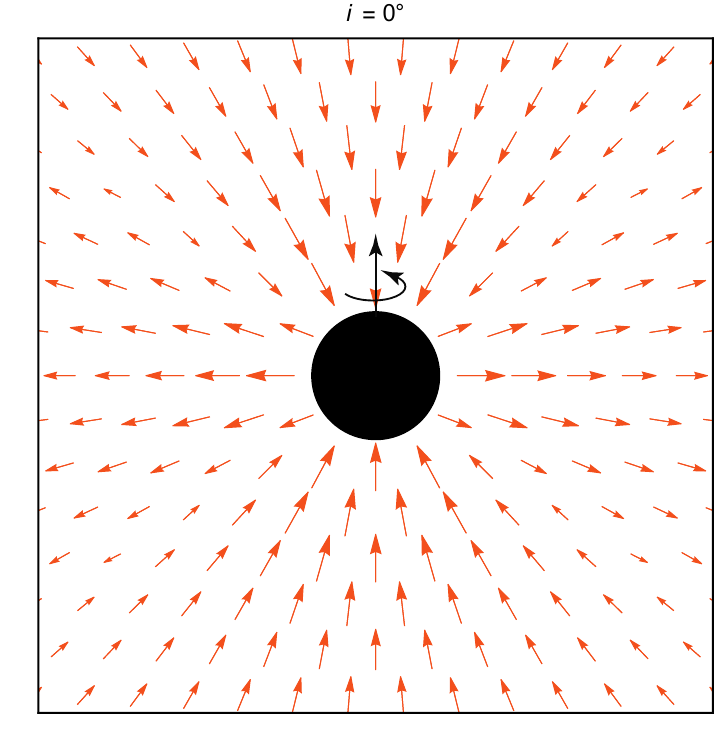}
\includegraphics[width=0.32\textwidth]{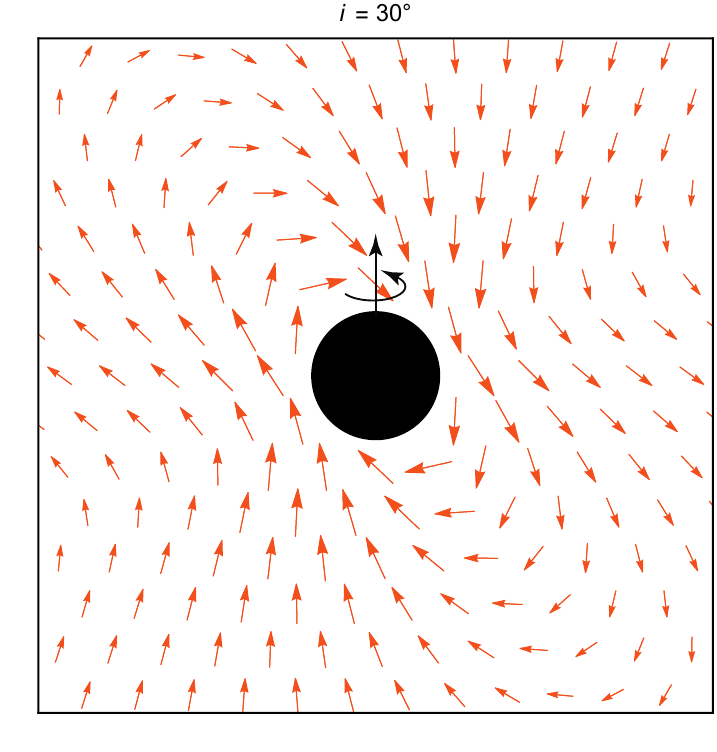}
\includegraphics[width=0.32\textwidth]{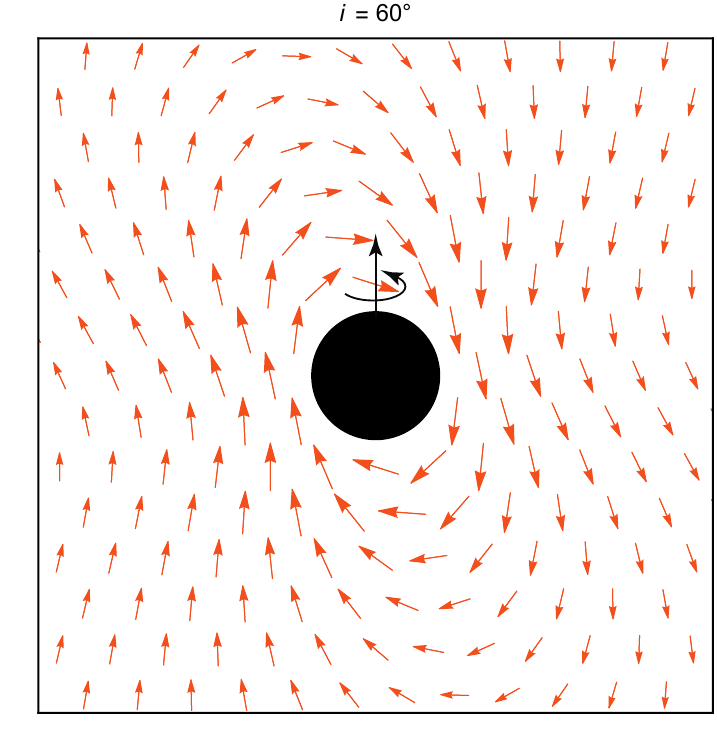}
\end{minipage}
%\hfill
\begin{minipage}[b]{\textwidth}
\centering
\includegraphics[width=0.32\textwidth]{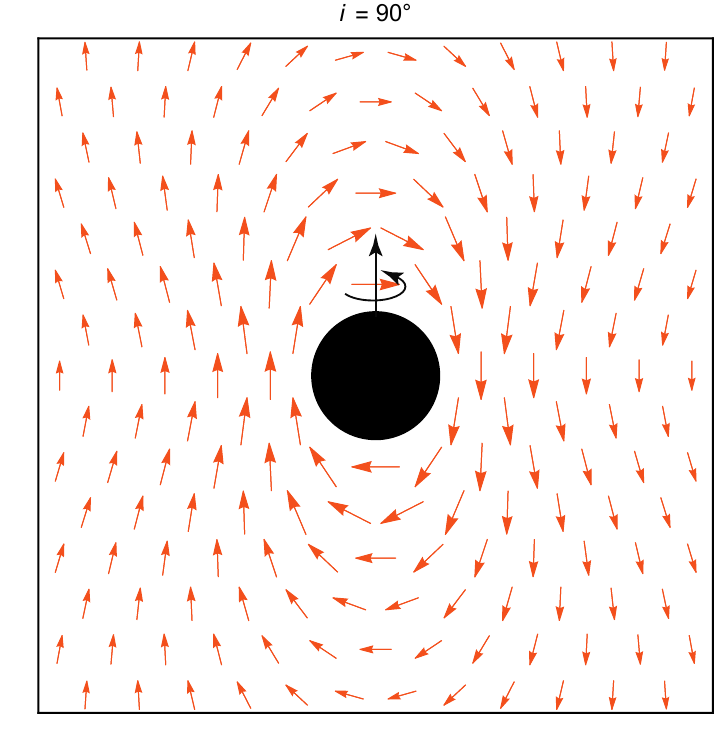}
\includegraphics[width=0.32\textwidth]{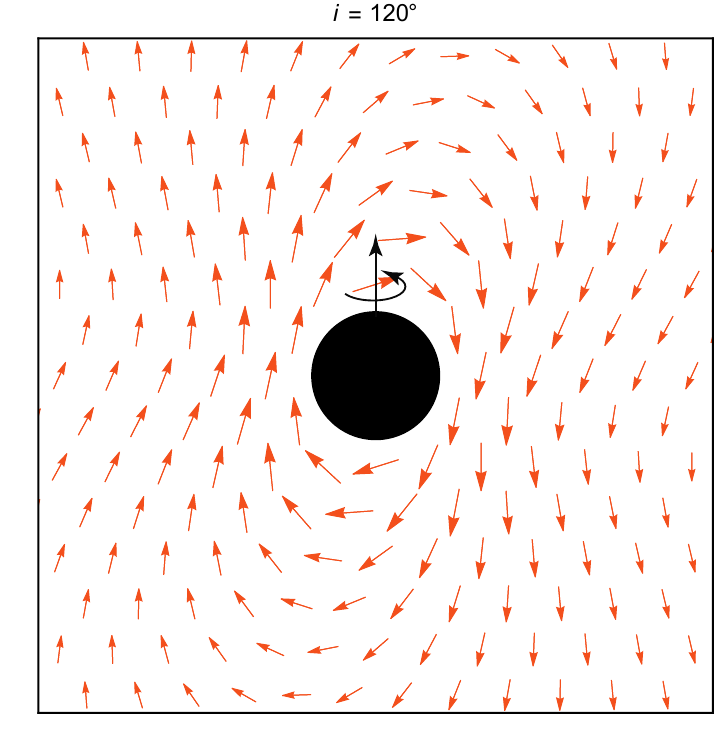}
\includegraphics[width=0.32\textwidth]{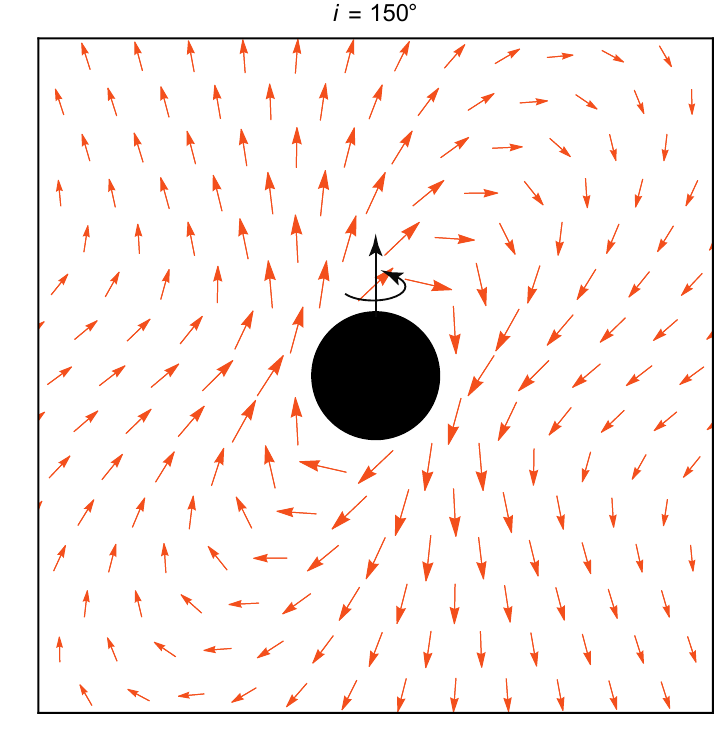}
\end{minipage}
\caption{Schematic representation of the electric fields measured by ZAMOs for selected inclinations $i$. The black disk denotes the BH, and its rotation axis is vertical. The arrow size is proportional to the field strength, which increases toward the horizon. Observe that the field is axisymmetric only for $i=0^{\circ}$ (or $i=180^{\circ}$).}
\label{fig:ZAMO_E}
\end{figure*}

The physical measurements are obtained by projecting $E^{\mu}$ and $B^{\mu}$ onto the observer's frame
\begin{align}
e_{0}^{}&=Z,\label{eq:ZAMOframe1}\\
e_{1}^{}&=\frac{1}{\sqrt{g_{rr}}}\frac{\partial}{\partial r},\\
e_{2}^{}&=\frac{1}{\sqrt{g_{\theta\theta}}}\frac{\partial}{\partial\theta},\\
e_{3}^{}&=\frac{1}{\sqrt{g_{\varphi\varphi}}}\frac{\partial}{\partial\varphi}.\label{eq:ZAMOframe2}
\end{align}
Specifically, the nonvanishing electric and magnetic components in this orthonormal basis are\footnote{Here we label the components in the ZAMO frame by a hat. Note that some authors put the hat over the indices, like $E_{\hat{1}}$, which means exactly the same thing.}
\begin{align}
\hat{E}_{1}&=-\frac{\sqrt{-g^{tt}}}{\sqrt{g_{rr}}}\left(F_{tr}-\omega F_{r\varphi}\right),\\
\hat{E}_{2}&=-\frac{\sqrt{-g^{tt}}}{\sqrt{g_{\theta\theta}}}\left(F_{t\theta}-\omega F_{\theta\varphi}\right),\\
\hat{E}_{3}&=-\frac{\sqrt{-g^{tt}}}{\sqrt{g_{\varphi\varphi}}}F_{t\varphi},\\
\hat{B}_{1}&=\frac{\sqrt{g_{rr}}}{\sqrt{-g}\sqrt{-g^{tt}}}F_{\theta\varphi},\\
\hat{B}_{2}&=-\frac{\sqrt{g_{\theta\theta}}}{\sqrt{-g}\sqrt{-g^{tt}}}F_{r\varphi},\\
\hat{B}_{3}&=\frac{\sqrt{g_{\varphi\varphi}}}{\sqrt{-g}\sqrt{-g^{tt}}}F_{r\theta}.
\end{align}
The full expressions are too cumbersome to be written explicitly. Instead, we present the electric fields for different inclinations in Fig.~\ref{fig:ZAMO_E} for intuitive insight. When $a=0$ or $B=0$, we have $\hat{E}_{1}=\hat{E}_{2}=\hat{E}_{3}=0$, which confirms that the electric field is induced by the BH rotation and the external magnetic field.
%%%%%%%%%%%%%%%%%%%%%%%%%%%%%%%%%%%%%%%%%%%%%%%%%%%%%%%%%%%%%%%%%%%%%%%%%%%%
\section{The dyadoregion}\label{sec:dyadoregion}
As mentioned in the introduction, the dyadoregion is a region where the electric field strength exceeds the critical value $E_{\text{c}}=m_{e}^{2}c^{3}/(e\hbar)$. However, it is known from the previous section that the electric field, and of course the magnetic field, are observer-dependent. By the laws of electrodynamics, in general only the following combinations of them are invariant:
\begin{align}
\mathcal{F}&\equiv\frac{1}{4}F_{\mu\nu}F^{\mu\nu}=\frac{1}{2}\left(\boldsymbol{B}^{2}-\boldsymbol{E}^{2}\right),\label{eq:scriptF}\\
\mathcal{G}&\equiv\frac{1}{4}F_{\mu\nu}{}^{\ast\!}F^{\mu\nu}=\boldsymbol{E}\cdot\boldsymbol{B}.\label{eq:scriptG}
\end{align}
Nevertheless, there exists a privileged observer for whom the measured electric and magnetic fields are parallel to each other. In that frame, the magnitudes of the parallel fields, called $\tilde{E}$ and $\tilde{B}$, are completely determined by the electromagnetic invariants $\mathcal{F}$ and $\mathcal{G}$ via
\begin{equation}\label{eq:tildeEtildeB}
\tilde{E}=\sqrt{\sqrt{\mathcal{F}^{2}+\mathcal{G}^{2}}-\mathcal{F}},\quad\tilde{B}=\sqrt{\sqrt{\mathcal{F}^{2}+\mathcal{G}^{2}}+\mathcal{F}}.
\end{equation}
This observer can be reached by performing a local Lorentz boost from the ZAMO frame \eqref{eq:ZAMOframe1}--\eqref{eq:ZAMOframe2} along the direction of $\boldsymbol{\hat{E}}\times\boldsymbol{\hat{B}}$ with the velocity \cite{LANDAU197543}
\begin{equation}
\frac{v}{1+v^{2}}=\frac{\lvert\boldsymbol{\hat{E}}\times\boldsymbol{\hat{B}}\rvert}{\hat{E}^{2}+\hat{B}^{2}},
\end{equation}
or explicitly
\begin{equation}
v=\frac{2\lvert\boldsymbol{\hat{E}}\times\boldsymbol{\hat{B}}\rvert}{\hat{E}^{2}+\hat{B}^{2}+\sqrt{(\hat{E}^{2}-\hat{B}^{2})^{2}+4(\boldsymbol{\hat{E}}\cdot\boldsymbol{\hat{B}})^{2}}},
\end{equation}
where $\hat{E}=\lvert\boldsymbol{\hat{E}}\rvert$ and $\hat{B}=\lvert\boldsymbol{\hat{B}}\rvert$. It is this observer that we will use to identify the dyadoregion.

Notably, Cherubini et al. \cite{Cherubini:2009ww} employed the Carter observer \cite{Carter:1968ks}, for the same purpose, to explore the vacuum polarization and $e^{+}e^{-}$ pair creation around a Kerr-Newman BH. In fact, as noted in \cite{Cherubini:2009ww}, the parallel-field observer constitutes the unique frame in which the flat-spacetime Schwinger treatment \cite{Schwinger:1951nm} can be applied locally, since an invariant description is intrinsically required for the rate of pair creation per unit four-volume,
\begin{align}
\frac{dN}{\sqrt{-g}d^{4}x}&=\frac{e^{2}\mathcal{G}}{4\pi^{2}\hbar^{2}}\sum_{l=1}^{\infty}\frac{1}{l}\coth\left(l\pi\sqrt{\frac{\sqrt{\mathcal{F}^{2}+\mathcal{G}^{2}}+\mathcal{F}}{\sqrt{\mathcal{F}^{2}+\mathcal{G}^{2}}-\mathcal{F}}}\right)\notag\\
&\qquad\times\exp\left(-l\pi\frac{E_{\text{c}}}{\sqrt{\sqrt{\mathcal{F}^{2}+\mathcal{G}^{2}}-\mathcal{F}}}\right).\label{eq:Schwingerrate}
\end{align}
By virtue of Eq.~\eqref{eq:tildeEtildeB}, the above equation simplifies to, in terms of the parallel electric and magnetic fields \cite{Damour:1974qv},
\begin{equation}\label{eq:rate}
\frac{dN}{\sqrt{-g}d^{4}x}=\frac{e^{2}\tilde{E}\tilde{B}}{4\pi^{2}\hbar^{2}}\sum_{l=1}^{\infty}\frac{1}{l}\coth\left(l\pi\frac{\tilde{B}}{\tilde{E}}\right)\exp\left(-l\pi\frac{E_{\text{c}}}{\tilde{E}}\right),
\end{equation}
a physically transparent expression which reveals that the pair creation is strongly suppressed by the exponential factor for $\tilde{E}\ll E_{\text{c}}$, and becomes significant when $\tilde{E}$ is of the same order as $E_{\text{c}}$. The creation rate goes to zero as $\tilde{E}\to 0$. In the limit $\tilde{B}\to 0$, however, one recovers the Schwinger formula for a pure electric field \cite{Schwinger:1951nm}
\begin{equation}
\frac{dN}{\sqrt{-g}d^{4}x}=\frac{1}{4\pi}\left(\frac{e\tilde{E}}{\pi\hbar}\right)^{\!\!2}\sum_{l=1}^{\infty}\frac{1}{l^{2}}\exp\left(-l\pi\frac{E_{\text{c}}}{\tilde{E}}\right).
\end{equation}
%%%%%%%%%%%%%%%%%%%%%%%%%%%%%%%%%%%%%%%%%%%%%%%%%%%%%%%%%%%%%%%%%%%%%%%%%%%%
\subsection{Dyadoregion morphology}
The exponential cutoff of the pair creation rate suggests that the dyadoregion surface should be defined by the field strength contour \cite{Cherubini:2009ww}
\begin{equation}\label{eq:contour}
\tilde{E}=kE_{\text{c}},
\end{equation}
where $k$ is a constant of order unity. To provide a quantitative comparison of our results with those of the aligned magnetic field case \cite{Cherubini:2025lnc}, here we choose to fix $k=1$ \footnote{For the field configuration under study, varying $k$ alters neither the dyadoregion topology nor any qualitative conclusions.}. In words, the dyadoregion is a locus of points outside the event horizon \eqref{eq:horizon} for which $\tilde{E}\geqslant E_{\text{c}}$. Recalling that the electromagnetic field we are considering lacks axial symmetry for generic inclinations, the dyadoregion is expected to be non-axisymmetric as well. That is to say, the dyadoregion radius $r_{\text{d}}=r_{\text{d}}(\theta,\varphi)$.

\begin{figure*}[htbp]
\centering
\begin{minipage}[b]{\textwidth}
\centering
\includegraphics[width=0.32\textwidth]{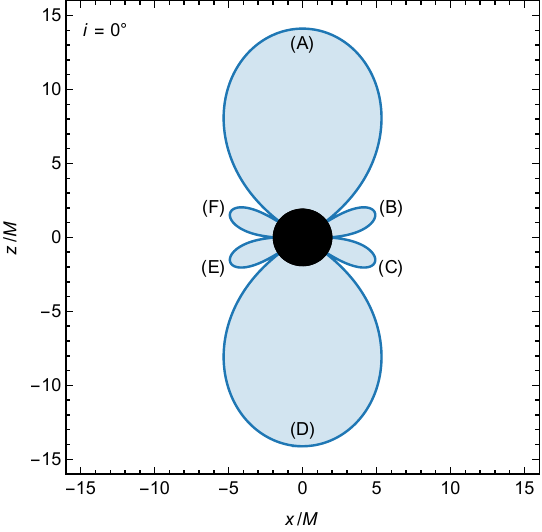}
\includegraphics[width=0.32\textwidth]{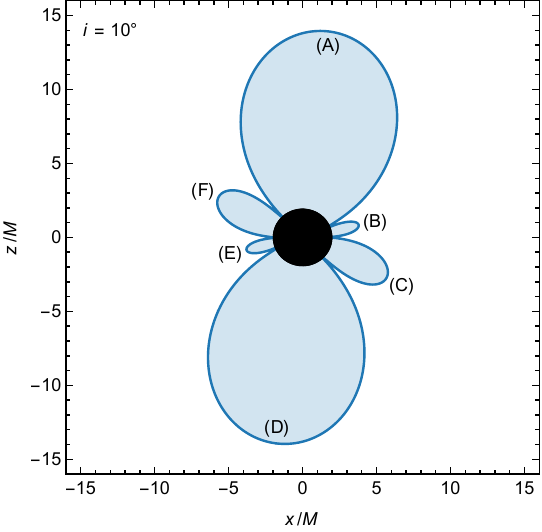}
\includegraphics[width=0.32\textwidth]{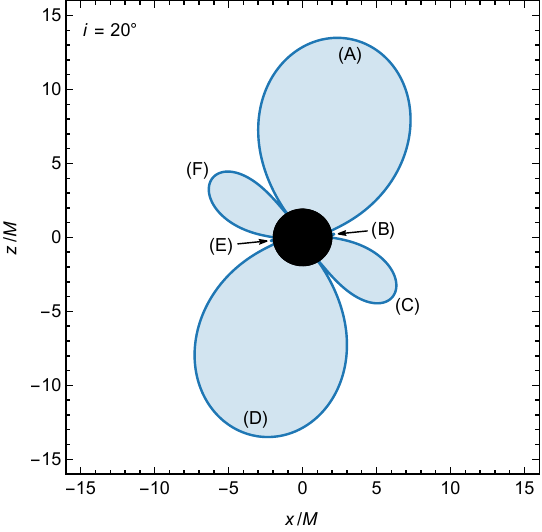}
\end{minipage}
%\hfill
\begin{minipage}[b]{\textwidth}
\centering
\includegraphics[width=0.32\textwidth]{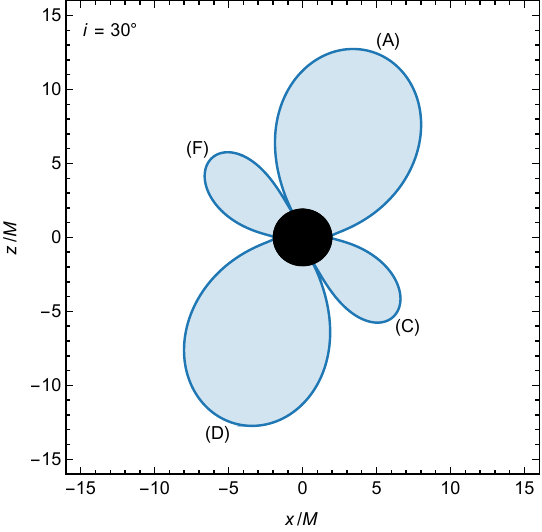}
\includegraphics[width=0.32\textwidth]{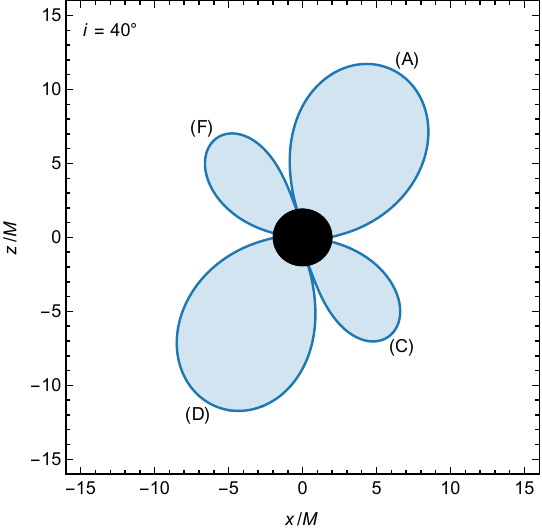}
\includegraphics[width=0.32\textwidth]{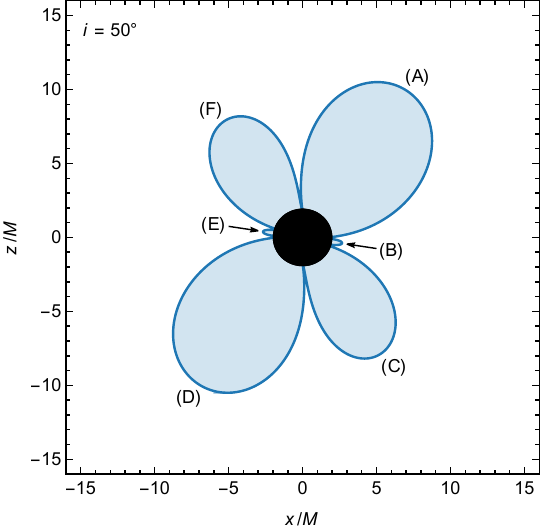}
\end{minipage}
%\hfill
\begin{minipage}[b]{\textwidth}
\centering
\includegraphics[width=0.32\textwidth]{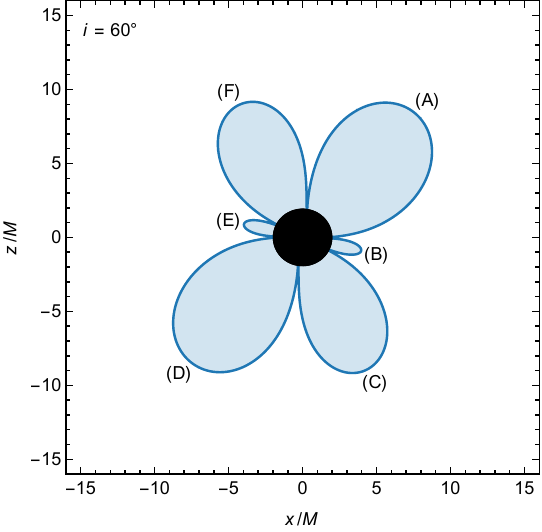}
\includegraphics[width=0.32\textwidth]{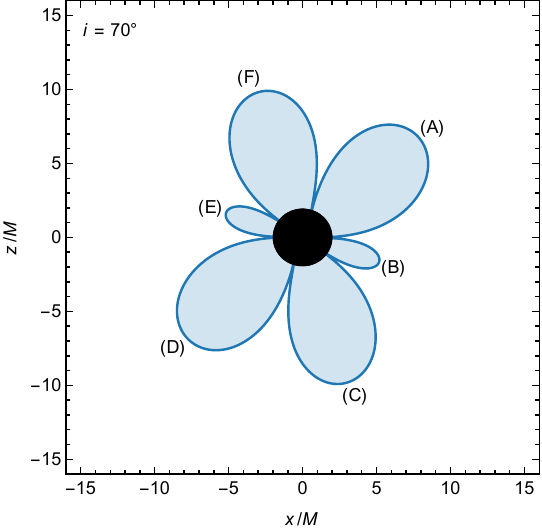}
\includegraphics[width=0.32\textwidth]{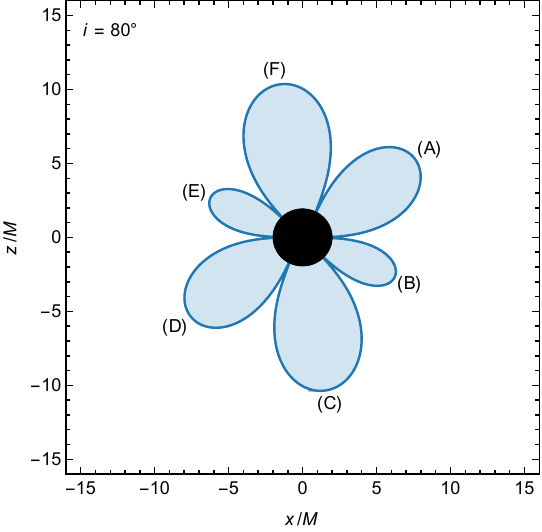}
\end{minipage}
%\hfill
\begin{minipage}[b]{\textwidth}
\centering
\includegraphics[width=0.32\textwidth]{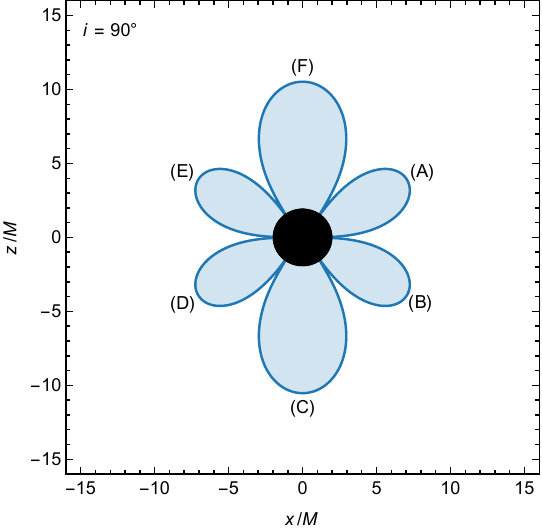}
\includegraphics[width=0.32\textwidth]{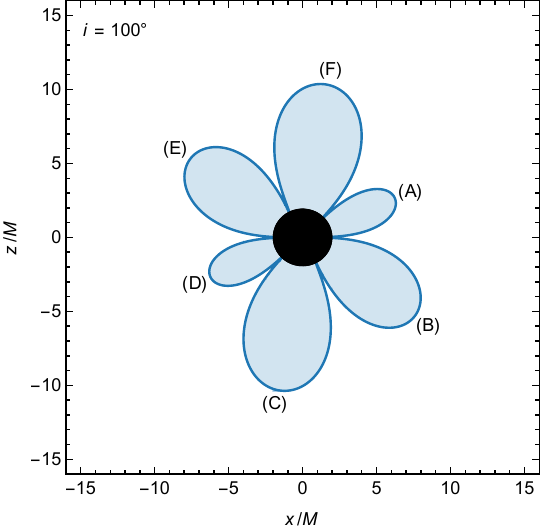}
\includegraphics[width=0.32\textwidth]{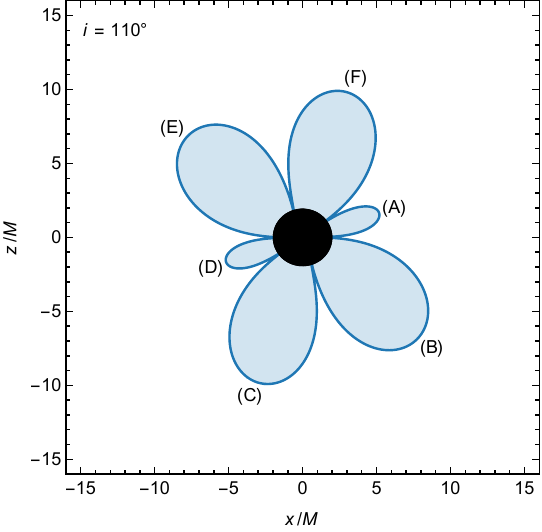}
\end{minipage}
\caption{The dyadoregion in the $x$-$z$ plane for different inclination angles $i$, with $\alpha=0.5$ and $\beta=200$. For these parameters, as $i$ increases from $0^{\circ}$ to $90^{\circ}$, the lobes (B) and (E) disappear at $i_{1}\approx 20.3^{\circ}$, and reappear at $i_{2}\approx 45.2^{\circ}$. The dyadoregion is axisymmetric if and only if $i=0^{\circ}$ (or $i=180^{\circ}$), which corresponds to the test field perfectly aligned with the BH spin axis. Moreover, the patterns for $i$ and $180^{\circ}-i$ are mirror symmetric with respect to the plane $x=0$.}
\label{fig:dyadoregion}
\end{figure*}

Fig.~\ref{fig:dyadoregion} shows the dyadoregion in the $x$-$z$ plane of the Kerr-Schild coordinates
\begin{align}
x&=(r\cos\phi-a\sin\phi)\sin\theta,\label{eq:KSx}\\
y&=(r\sin\phi+a\cos\phi)\sin\theta,\label{eq:KSy}\\
z&=r\cos\theta,\label{eq:KSz}
\end{align}
where $\phi$ is given by Eq.~\eqref{eq:phi}. In each subplot, we take the dimensionless spin $\alpha\equiv a/M=0.5$ and the magnetic field strength $\beta\equiv B/E_{\text{c}}=200$, which correspond to $B\approx 8.83\times 10^{15}\operatorname{G}$. It is evident from the figure that vacuum breakdown occurs in several lobes around the BH. Sometimes there are six lobes, and sometimes four. To better trace how these lobes evolve with the inclination angle $i$ in the projection plane, we mark them in a clockwise fashion as (A), (B), (C), (D), (E), and (F). The observed pattern is periodic with a period of $180^{\circ}$ and exhibits an alternating dominance among the lobes. Initially, at $i=0^{\circ}$, the lobes (A) and (D) on the polar axis are dominant, while the other four small lobes are of equal size and symmetric about the equatorial plane. As $i$ increases from $0^{\circ}$ to $90^{\circ}$, the (A,D) pair shrink; the (B,E) pair first shrink, then disappear at a critical angle $i_{1}$, reappear at another critical angle $i_{2}$, and then expand; while the (C,F) pair grow and eventually dominate. For $i\in(90^{\circ},180^{\circ}]$, (A,D) continue to shrink, vanish at $180^{\circ}-i_{2}$, reappear at $180^{\circ}-i_{1}$, and then expand; (B,E) gradually grow to dominate; and (C,F) shrink.

An analytic expression for the dyadoregion surface may be approximately obtained by performing a series expansion of the electric field strength $\tilde{E}$ in \eqref{eq:tildeEtildeB} at large $r$. This yields, to leading order,
\begin{equation}\label{eq:tildeE}
\tilde{E}\approx\frac{\alpha B}{\bar{r}^{2}}\lvert\cos\theta\rvert\lvert3(\cos i\cos\theta+\sin i\sin\theta\cos\varphi)^{2}-1\rvert,
\end{equation}
where $\bar{r}\equiv r/M$. Then, it follows from Eq.~\eqref{eq:contour} that
\begin{equation}\label{eq:barrd}
\bar{r}_{\text{d}}\approx\sqrt{\alpha\beta\lvert\cos\theta\rvert\lvert3(\cos i\cos\theta+\sin i\sin\theta\cos\varphi)^{2}-1\rvert},
\end{equation}
which is very helpful for quickly developing a 3D visualization of the dyadoregion with sufficient accuracy. For instance, when the magnetic field is exactly perpendicular to the BH spin axis, i.e., $i=90^{\circ}$ (or $270^{\circ}$), one might be misled by the bottom left panel of Fig.~\ref{fig:dyadoregion} into thinking that the dyadoregion becomes axisymmetric at this special inclination. However, a glance at Eq.~\eqref{eq:barrd} confirms that for $i=90^{\circ}$,
\begin{equation}
\bar{r}_{\text{d}}\approx\sqrt{\alpha\beta\lvert\cos\theta\rvert\lvert3\sin^{2}\theta\cos^{2}\varphi-1\rvert},
\end{equation}
which possesses only reflection symmetries with respect to the three coordinate planes.

\subsection{Minimum magnetic field for pair creation}
Given a spin $\alpha$ and an inclination $i$, not every magnetic field induces vacuum breakdown around the BH. Roughly speaking, if $\beta$ is too small, the entire dyadoregion could be hidden behind the horizon. We can obtain the minimum magnetic field strength $\beta_{\text{min}}$ required to initiate pair creation by demanding that the maximum of $\tilde{E}$ equals the critical value $E_{\text{c}}$, i.e.,
\begin{equation}
\max\tilde{E}(\alpha,\beta_{\text{min}},i;r,\theta,\varphi)=E_{\text{c}},
\end{equation}
where $r\geqslant r_{+}$.

\begin{figure}[htbp]
\includegraphics[width=0.475\textwidth]{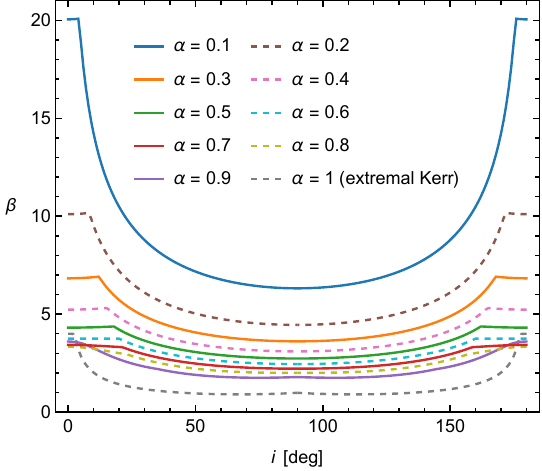}
\caption{The minimum magnetic field $\beta_{\text{min}}$ versus the inclination $i$, shown for ten selected BH spins $\alpha$. The configurations above each $\alpha$-line are allowed for pair creation. A key feature is that vacuum breakdown occurs more easily for most inclinations, especially near $90^{\circ}$. However, for $\alpha\lesssim 0.6$ , there are four angles over the full range of $i$ where the breakdown becomes more difficult. For example, for $\alpha=0.5$ (green solid line), they are $18^{\circ}$, $162^{\circ}$, $198^{\circ}$, and $342^{\circ}$, where $\beta_{\text{min}}\approx 4.37$, corresponding to $1.93\times 10^{14}\operatorname{G}$. Note that the value $\beta=200$ we adopted in Fig.~\ref{fig:dyadoregion} is far above this threshold.}
\label{fig:betamin}
\end{figure}

In the aligned case ($i=0^{\circ}$ or $180^{\circ}$), $\tilde{E}$ is independent of the azimuthal coordinate $\varphi$ and its maximum is attained on the polar axis ($\theta=0$). The result is simply \cite{Cherubini:2025lnc}
\begin{equation}\label{eq:betamin4aligned}
\beta_{\text{min}}=
\begin{cases}
\dfrac{1}{\alpha}\left(1+\dfrac{1}{\sqrt{1-\alpha^{2}}}\right),&0<\alpha\leqslant\dfrac{\sqrt{3}}{2},\\
4\alpha,&\dfrac{\sqrt{3}}{2}\leqslant\alpha\leqslant 1.
\end{cases}
\end{equation}
For a generic inclination, however, we must maximize the function $\tilde{E}$ over the region $r\geqslant r_{+}$ for each combination of $\alpha$ and $i$. We could not derive an analytic formula for the misaligned case, but the numerical evaluation is presented in Fig.~\ref{fig:betamin}. Compared with the perfectly aligned case, vacuum breakdown is more easily triggered for a wide range of inclinations, especially for the perpendicular configuration ($i=90^{\circ}, 270^{\circ}$) or nearly perpendicular ones. Therefore, a misaligned magnetic field generally favors pair creation around the BH more than the aligned field. This is because the loss of axial symmetry enables the induced electric field to be focused in specific directions, leading to a stronger peak field $\tilde{E}_{\text{max}}$ that more readily exceeds the critical value.

Another intriguing fact is that an extremal Kerr BH allows pair creation by vacuum breakdown even when the magnetic field is aligned with its spin axis. Indeed, for $\alpha=1$ and $i=0^{\circ}$, the maximum of $\tilde{E}$ is located at $r=\sqrt{3}M$, and Eq.~\eqref{eq:betamin4aligned} gives $\beta_{\text{min}}=4$ (see the two ends of the gray dashed line in Fig.~\ref{fig:betamin}), or approximately $1.77\times 10^{14}$ in Gauss. This threshold was not reported by Cherubini et al. in Ref.~\cite{Cherubini:2025lnc}. Such a finding reveals that vacuum breakdown is not completely suppressed by the BH Meissner effect \cite{Bicak1985}, which asserts that a maximally rotating BH expels out any stationary and axisymmetric external electromagnetic field from the event horizon, like a superconductor. Misaligned configurations do not experience the Meissner effect \cite{Bicak1985,Dovciak_2000}.
%%%%%%%%%%%%%%%%%%%%%%%%%%%%%%%%%%%%%%%%%%%%%%%%%%%%%%%%%%%%%%%%%%%%%%%%%%%%
\subsection{Electromagnetic energy in the dyadoregion}
Now we turn to evaluating the electromagnetic energy available for pair creation. In Kerr spacetime, the energy of an electromagnetic field $F_{\mu\nu}$ on a constant-$t$ hypersurface $\Sigma$ can be defined by the Killing integral \cite{Cherubini:2009ww}
\begin{equation}\label{eq:Killinginte}
\mathcal{E}=\int_{\Sigma}T_{\mu\nu}\xi^{\mu}n^{\nu},
\end{equation}
where $\xi^{\mu}$ is the timelike Killing vector as in Eq.~\eqref{eq:Killing}, $n^{\nu}$ is the unit normal to the hypersurface $\Sigma$ at each event, and the stress-energy tensor is of course
\begin{equation}\label{eq:Tmunu}
T_{\mu\nu}=\frac{1}{4\pi}\left(F_{\mu\sigma}F_{\nu}{}^{\sigma}-\frac{1}{4}g_{\mu\nu}F_{\sigma\rho}F^{\sigma\rho}\right).
\end{equation}
By virtue of the Killing equation and the Einstein field equation, it is easy to see that the current $\mathcal{J}^{\mu}=T^{\mu\nu}\xi_{\nu}$ is covariantly conserved. Hence, the integral \eqref{eq:Killinginte} does not depend on the choice of time slice. Note also that the ZAMO four-velocity \eqref{eq:ZAMO} is orthogonal to the hypersurfaces of constant $t$; the normal vector $n$ is exactly $Z$.

\begin{figure*}[htbp]
\centering
\begin{minipage}[b]{\textwidth}
\centering
\includegraphics[width=0.475\textwidth]{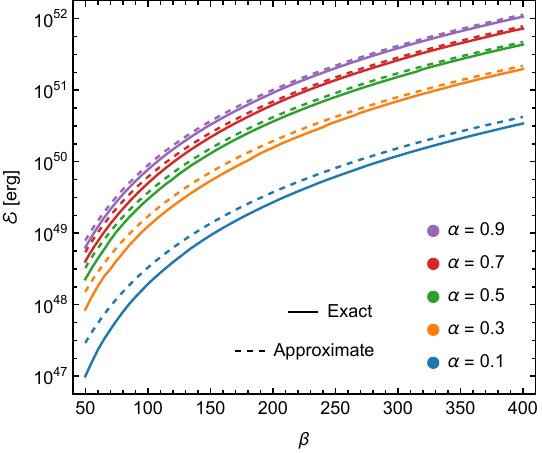}\hfill
\includegraphics[width=0.4839\textwidth]{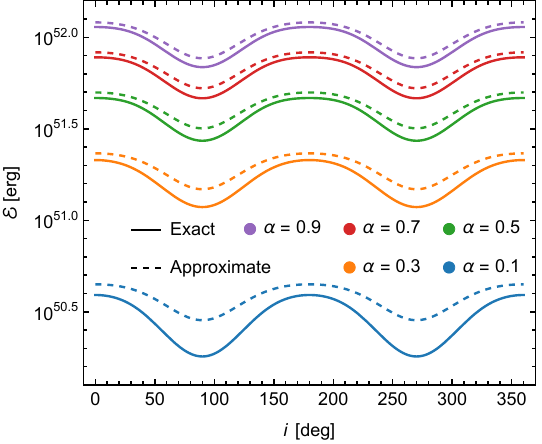}
\end{minipage}
\caption{Left: Dyadoregion electromagnetic energy $\mathcal{E}$ as a function of the magnetic field strength $\beta\in[50,400]$, with $i=30^{\circ}$. Right: Dependence of the energy $\mathcal{E}$ on the magnetic field inclination $i$ for $\beta=400$. In both panels, colors are used to identify different spins $\alpha$; solid lines represent the numerical results obtained from Eq.~\eqref{eq:energyEM}, whereas dashed lines correspond to the approximate values given by Eq.~\eqref{eq:scalinglaw}. The BH mass is set to $M=3M_{\odot}$.}
\label{fig:energyEM}
\end{figure*}

As $e^{+}e^{-}$ pairs are mainly produced within the dyadoregion, we need not include the total electromagnetic energy over the entire space $\Sigma$. For our purpose, it is sufficient to restrict the integral \eqref{eq:Killinginte} to the dyadoregion. This yields a consistent quasilocal energy
\begin{equation}
\mathcal{E}=\int_{0}^{2\pi}\!\!\int_{0}^{\pi}\!\int_{r_{+}}^{r_{\text{d}}(\theta,\varphi)}\!\!\sqrt{-g^{tt}}\left(T_{tt}+\omega T_{t\varphi}\right)d\Sigma,
\end{equation}
where Eqs.~\eqref{eq:Killing} and \eqref{eq:ZAMO} are used, $\omega=-g_{t\varphi}/g_{\varphi\varphi}$, and the induced volume element $d\Sigma=\sqrt{g_{rr}g_{\theta\theta}g_{\varphi\varphi}}drd\theta d\varphi$. With the help of Eq.~\eqref{eq:Tmunu}, the integral can be performed now. Alternatively, one can invert the orthonormal basis \eqref{eq:ZAMOframe1}--\eqref{eq:ZAMOframe2} to translate the coordinate components $T_{tt}$ and $T_{t\varphi}$ into the physical quantities in the ZAMO frame. After feeding those metric functions, we arrive at
\begin{equation}\label{eq:energyEM}
\mathcal{E}=\int_{0}^{2\pi}\!\!\int_{0}^{\pi}\!\int_{r_{+}}^{r_{\text{d}}}\!\left(U+\frac{2Mar}{\rho^{2}\sqrt{\Delta}}\sin\theta\hat{S}_{3}\right)\rho^{2}\sin\theta drd\theta d\varphi,
\end{equation}
in which
\begin{equation}
U=\frac{\hat{E}^{2}+\hat{B}^{2}}{8\pi},\quad\hat{S}_{3}=\frac{(\boldsymbol{\hat{E}}\times\boldsymbol{\hat{B}})_{3}}{4\pi}
\end{equation}
are the energy density and the toroidal Poynting flux measured by the ZAMO, respectively.

Since $U$ and $\hat{S}_{3}$ have no axisymmetry in general, to evaluate the integral \eqref{eq:energyEM} for a given triple ($\alpha,\beta,i$) we numerically solve the equation \eqref{eq:contour} in each ($\theta,\varphi$) direction to obtain the dyadoregion surface $r_{\text{d}}(\alpha,\beta,i;\theta,\varphi)$, which serves as the upper limit of the integration. As shown in Fig.~\ref{fig:energyEM}, for a fixed inclination $i$, the electromagnetic energy $\mathcal{E}$ stored in the dyadoregion increases monotonically with both the BH spin $\alpha$ and the field strength $\beta$. For fixed $\alpha$ and $\beta$, $\mathcal{E}$ oscillates with $i$ in a cosine-like manner, with a period of $\pi$. The energy $\mathcal{E}$ is maximal when the magnetic field is parallel to the BH spin axis and minimal when it is perpendicular, with the maximum variation within one order of magnitude.

A simple but reasonably accurate analytic expression for the dyadoregion energy can be established as follows. Let us first observe that at large $r$,
\begin{align}
\hat{E}&\approx\frac{\alpha B}{\bar{r}^{2}}\left\{\left[\cos i(3\cos^{2}\theta-1)+3\sin i\sin\theta\cos\theta\cos\varphi\right]^{2}\right.\notag\\
&\left.\phantom{\left[(\cos^{2}\theta)\right]^{2}}+\sin^{2}i(1-\sin^{2}\theta\sin^{2}\varphi)\right\}^{1/2},\\
\hat{B}&=\tilde{B}\approx B\left\{1-\frac{1}{\bar{r}}\left[(\cos i\sin\theta-\sin i\cos\theta\cos\varphi)^{2}\right.\right.\notag\\
&\qquad\qquad\qquad\qquad\left.\phantom{\frac{1}{\bar{r}}}\left.+\,\sin^{2}i\sin^{2}\varphi\right]\right\}.\label{eq:tildeB}
\end{align}
This means that $\hat{B}\gg\hat{E}$ and the electromagnetic energy is magnetically dominated, i.e.,
\begin{equation}
U\approx\frac{\hat{B}^{2}}{8\pi}.
\end{equation}
On the other hand, as $r\to\infty$,
\begin{equation}
\frac{2Mar}{\rho^{2}\sqrt{\Delta}}\approx\frac{2\alpha}{\bar{r}^{2}},\quad\hat{S}_{3}\approx\frac{\alpha B^{2}}{8\pi\bar{r}^{2}}\mathcal{A}(i;\theta,\varphi),
\end{equation}
where $\mathcal{A}$ is an angular function of order unity. Collecting these results and keeping only terms up to the first order in $\alpha$, the integral \eqref{eq:energyEM} approximates to
\begin{align}
\mathcal{E}&\approx\frac{1}{8\pi}\int_{0}^{2\pi}\!\!\int_{0}^{\pi}\!\int_{r_{+}}^{r_{\text{d}}}\hat{B}^{2}r^{2}\sin\theta drd\theta d\varphi\notag\\
&\approx\frac{M^{3}B^{2}}{24\pi}\int_{0}^{2\pi}\!\!\int_{0}^{\pi}\bar{r}_{\text{d}}^{3}\sin\theta d\theta d\varphi\notag\\
&\approx 5.23\times 10^{41}\mu^{3}\alpha^{3/2}\beta^{7/2}\eta(i)\operatorname{erg},\label{eq:scalinglaw}
\end{align}
where Eq.~\eqref{eq:barrd} is used in the last step, $\mu\equiv M/M_{\odot}$, and
\begin{align}
\eta(i)&\equiv\frac{1}{2\pi}\int_{0}^{2\pi}\!\!d\varphi\int_{0}^{\pi}d\theta\sin\theta\lvert\cos\theta\rvert^{3/2}\notag\\
&\qquad\times\lvert3(\cos i\cos\theta+\sin i\sin\theta\cos\varphi)^{2}-1\rvert^{3/2}\notag\\
&\approx 0.658+0.142\cos(2i)-0.018\cos(4i).\label{eq:etaofi}
\end{align}
When the magnetic field is aligned with the BH spin, we have $\eta(0)\approx 0.782$, which agrees with the result in Ref.~\cite{Cherubini:2025lnc}. This analytic approximation becomes accurate for high spin, strong magnetic field, and small inclination, as can be seen in Fig.~\ref{fig:energyEM}. For instance, for $\mu=3$, $\alpha=0.9$, $\beta=400$, and $i=30^{\circ}$, the energy calculated from Eq.~\eqref{eq:energyEM} is about $1.06\times 10^{52}\operatorname{erg}$, while Eq.~\eqref{eq:scalinglaw} predicts $1.14\times 10^{52}\operatorname{erg}$, with a relative difference of $7.5\%$.

We have thus fully characterized the electromagnetic energy in the dyadoregion. The dyadoregion energy $\mathcal{E}$ provides the total energy budget for the pair creation by vacuum breakdown, as well as for the subsequent plasma dynamics and its eventual radiation. It is evident from Fig.~\ref{fig:dyadoregion} that the energy distribution is not isotropic; instead, depending on the inclination $i$, it is confined to several lobes of different sizes and orientations.

However, astronomers typically interpret the received flux under the assumption of isotropic emission, which would overestimate the intrinsic energy of the dyadoregion. We conclude this section by deriving a beaming factor $f_{b}(i)$ that measures the degree of collimation and allows a conversion between the actual dyadoregion energy $\mathcal{E}$ and the observed isotropic energy $\mathcal{E}_{\text{iso}}$ via
\begin{equation}\label{eq:Eiso}
\mathcal{E}=f_{b}(i)\mathcal{E}_{\text{iso}}.
\end{equation}

In the aligned geometry, the dyadoregion energy is mostly concentrated in two polar lobes of equal size (see the top left panel of Fig.~\ref{fig:dyadoregion}). By treating the dyadoregion energy as the magnetic energy contained in two opposite spherical cones with apex at the BH center and directed along the polar axis, Cherubini et al. \cite{Cherubini:2025lnc} found a beaming factor of $\eta(0)/2^{5/2}\approx 0.14$, which corresponds to a semi-aperture angle $\theta_{b}\approx 30.42^{\circ}$ for each cone. This implies that, for the aligned case, the actual dyadoregion energy is only about $14\%$ of the isotropic energy, or conversely, the observed $\mathcal{E}_{\text{iso}}$ is about $7.14$ times the intrinsic energy $\mathcal{E}$, based on Eq.~\eqref{eq:Eiso}.

For the misaligned configuration, however, the situation is slightly more complicated. There are two reasons: first, the energy distribution varies with the inclination $i$ and has no fixed dominant orientation; second, the fractional energy contributed by different lobes is not constant, especially for the perpendicular or nearly perpendicular configurations, where no lobe can be a priori neglected. Consequently, we cannot simply capture the misaligned dyadoregion energy with the twin-cone model. A more general equivalent scheme is needed.

Here we adopt the $90\%$ energy containment solid angle, denoted by $\Omega_{90}(i)$. To be specific, for the angular energy distribution under investigation
\begin{align}
\frac{d\mathcal{E}}{d\Omega}&\approx\frac{M^{3}B^{2}}{24\pi}\bar{r}_{\text{d}}^{3}=\frac{M^{3}B^{2}}{24\pi}\alpha^{3/2}\beta^{3/2}\lvert\cos\theta\rvert^{3/2}\notag\\
&\qquad\times\lvert3(\cos i\cos\theta+\sin i\sin\theta\cos\varphi)^{2}-1\rvert^{3/2},\label{eq:energydensity}
\end{align}
we identify a set of preferred angular regions (which may consist of several disconnected patches) such that the total solid angle subtended by these regions is minimized while enclosing $90\%$ of the total energy $\mathcal{E}$. To this end, we first divide the sphere into a $200\times 200$ grid. For each inclination $i$, we compute the energy density per cell using Eq.~\eqref{eq:energydensity}, then accumulate the cell energies in descending order of energy density until $90\%$ of the total energy is reached. The solid angle covered by the selected cells in this process is taken as $\Omega_{90}(i)$. Accordingly, the desired beaming factor is
\begin{equation}\label{eq:deffbi}
f_{b}(i)=\frac{\Omega_{90}(i)}{4\pi}.
\end{equation}
Note that the equivalent solid angle and thus the beaming factor depend only on the inclination $i$, since the common prefactor $M^{3}B^{2}\alpha^{3/2}\beta^{3/2}/24\pi$ in Eq.~\eqref{eq:energydensity} cancels when determining the energy fraction.

\begin{figure}[htbp]
\includegraphics[width=0.475\textwidth]{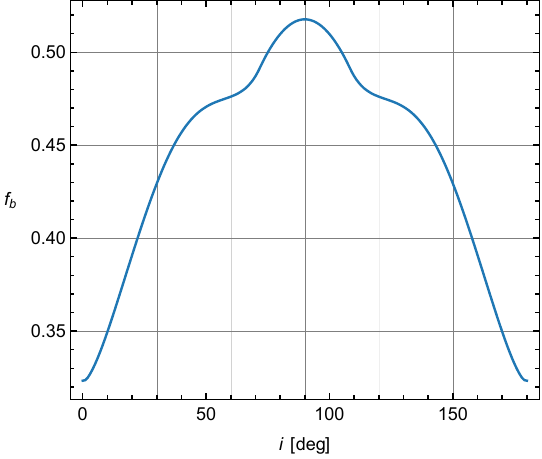}
\caption{Plot of the beaming factor $f_{b}(i)$ defined by Eq.~\eqref{eq:deffbi}. The curve exhibits a symmetric bell-shaped profile. In the rising branch, three inflection points are observed at $i\approx 17^{\circ}$, $58^{\circ}$, and $72^{\circ}$. While the first one is merely a minor inflection, the latter two are physically relevant. The inflection at $58^{\circ}$ marks the transition of the dyadoregion energy distribution from a two-lobe-dominated pattern to a four-lobe-dominated one, whereas the inflection at $72^{\circ}$ signals the transition back to a two-lobe-dominated configuration.}
\label{fig:fbofi}
\end{figure}

We plot the function $f_{b}=f_{b}(i)$ in Fig.~\ref{fig:fbofi}. The beaming factor follows a bell-shaped curve, exhibiting a minimum of about $0.323$ at $i=0^{\circ}$ or $180^{\circ}$ and a maximum of about $0.518$ at $i=90^{\circ}$. This shows that the dyadoregion energy $\mathcal{E}$ is most concentrated in the aligned case and most dispersed (closer to isotropic) in the perpendicular one, which is consistent with the features depicted in Fig.~\ref{fig:dyadoregion}. The relative increase in $f_{b}(i)$ from alignment to perpendicular amounts to $60\%$. It should be noted that our beaming factor for $i=0^{\circ}$ is approximately $2.3$ times larger than that of Cherubini et al. \cite{Cherubini:2025lnc}, due to the different definition employed.

Given an observed isotropic energy $\mathcal{E}_{\text{iso}}$, if the magnetic field inclination $i$ is measured independently (e.g., through polarization patterns or jet morphology), the intrinsic dyadoregion energy $\mathcal{E}$ can be inferred via Eq.~\eqref{eq:Eiso}. Otherwise, the uncertainty in the inclination leads to a systematic error of at most $\pm 23\%$ in the estimated energy when the mid-value of $0.4205$ is used. Compared to the much larger uncertainties associated with other model parameters (such as the BH spin $\alpha$), the error introduced by the unknown inclination $i$ is relatively small. Therefore, the estimate of $\mathcal{E}$ is robust against inclination uncertainties to an acceptable level. Conversely, it is also possible to predict the observed isotropic energy from a theoretical value. For example, with $\mu=3$, $\alpha=0.9$, $\beta=400$, and $i=30^{\circ}$, Eq.~\eqref{eq:scalinglaw} yields $\mathcal{E}\approx 1.14\times 10^{52}\operatorname{erg}$, so we have $\mathcal{E}_{\text{iso}}=\mathcal{E}/f_{b}(30^{\circ})\approx 2.66\times 10^{52}\operatorname{erg}$.

%%%%%%%%%%%%%%%%%%%%%%%%%%%%%%%%%%%%%%%%%%%%%%%%%%%%%%%%%%%%%%%%%%%%%%%%%%%%
\section{Pair plasma thermodynamics}\label{sec:thermo}
Inside the dyadoregion, the $e^{+}e^{-}$ pairs created by vacuum breakdown quickly thermalize via collisions and radiation processes on a very short timescale \cite{Aksenov:0227,Aksenov:2009dy}. Consequently, the BH's surroundings are filled with a thermalized $e^{+}e^{-}\gamma$ plasma in local thermodynamic equilibrium, described by a single temperature $T$. After that, the pair-photon plasma undergoes rapid outward expansion and adiabatic cooling, eventually becoming transparent and releasing the trapped radiation. This is the origin of the prompt emission of GRBs.

The evolution of the plasma is governed by the general relativistic magnetohydrodynamic (GRMHD) equations. In the following, we determine the thermodynamic properties of the pair plasma immediately after thermalization, which provide the initial conditions for the subsequent dynamical evolution of the plasma.

In thermal equilibrium, electrons and positrons obey Fermi-Dirac statistics. With zero chemical potential, the proper number density of $e^{+}e^{-}$ pairs, denoted by $n$, is related to the temperature $T$ by
\begin{equation}
n=\frac{t^{3}}{\pi^{2}\lambdabar_{e}^{3}}\int_{0}^{\infty}\!\frac{y^{2}dy}{e^{\sqrt{y^{2}+1/t^{2}}}+1},
\end{equation}
where $t\equiv k_{B}T/(m_{e}c^{2})$ and $\lambdabar_{e}=\hbar/(m_{e}c)$ is the reduced Compton wavelength. In the ultrarelativistic limit $t\gg 1$, the integral is well approximated by
\begin{equation}\label{eq:nofT}
n\approx\frac{3\zeta(3)}{2\pi^{2}\lambdabar_{e}^{3}}\left(\frac{k_{B}T}{m_{e}c^{2}}\right)^{\!\!3},
\end{equation}
with $\zeta(3)\approx 1.202$. On the other hand, keeping only the $l=1$ term in the invariant pair creation rate \eqref{eq:rate} and restoring the speed of light $c$, we have
\begin{equation}
\frac{dN}{\sqrt{-g}d^{4}x}\approx\frac{e^{2}\tilde{E}\tilde{B}}{4\pi^{2}\hbar^{2}c}\coth\left(\pi\frac{\tilde{B}}{\tilde{E}}\right)\exp\left(-\pi\frac{E_{\text{c}}}{\tilde{E}}\right).
\end{equation}
In the local rest frame of the plasma, the invariant four-volume element $\sqrt{-g}d^{4}x$ can be written as $d\tau dV$, with $d\tau$ the proper time and $dV$ the proper volume measured in that frame. Taking $d\tau$ as the Compton time $\hbar/(m_{e}c^{2})$ and using $\tilde{B}\gg\tilde{E}$, the local proper number density $n$ in terms of the parallel fields is then estimated as
\begin{equation}\label{eq:nofEandB}
n=\frac{dN}{dV}\approx\frac{e^{2}}{4\pi^{2}\hbar m_{e}c^{3}}\tilde{E}\tilde{B}\exp\left(-\pi\frac{E_{\text{c}}}{\tilde{E}}\right).
\end{equation}
By comparing Eq.~\eqref{eq:nofT} with Eq.~\eqref{eq:nofEandB}, the equilibrium temperature is found to be
\begin{equation}\label{eq:kBT}
k_{B}T=\frac{m_{e}c^{2}}{[6\zeta(3)]^{1/3}}\left(\frac{\tilde{E}\tilde{B}}{E_{\text{c}}^{2}}\right)^{\!\!1/3}\!\!\exp\left(-\frac{\pi}{3}\frac{E_{\text{c}}}{\tilde{E}}\right).
\end{equation}
According to Eq.~\eqref{eq:notaxisymmetric}, the electromagnetic field is not axisymmetric in general. Hence, the temperature depends on all three spherical coordinates, i.e., $T=T(r,\theta,\varphi)$, and so does the pair density $n$. Once the temperature is known, the energy density and pressure of the $e^{+}e^{-}\gamma$ plasma follow from the standard expressions for a relativistic Fermi-Bose mixture:
\begin{align}
\epsilon&=aT^{4}\left(1+\frac{30}{\pi^{4}}\int_{0}^{\infty}\frac{y^{2}\sqrt{y^{2}+1/t^{2}}dy}{e^{\sqrt{y^{2}+1/t^{2}}}+1}\right)\approx\frac{11}{4}aT^{4},\\
P&=\frac{1}{3}aT^{4}\left(1+\frac{30}{\pi^{4}}\int_{0}^{\infty}\frac{y^{4}/\sqrt{y^{2}+1/t^{2}}dy}{e^{\sqrt{y^{2}+1/t^{2}}}+1}\right)\approx\frac{1}{3}\epsilon,\label{eq:pressure}
\end{align}
where $a=\pi^{2}k_{B}^{4}/(15\hbar^{3}c^{3})$ is the radiation constant.

\begin{figure*}[htbp]
\centering
\begin{minipage}[b]{\textwidth}
\centering
\includegraphics[width=0.475\textwidth]{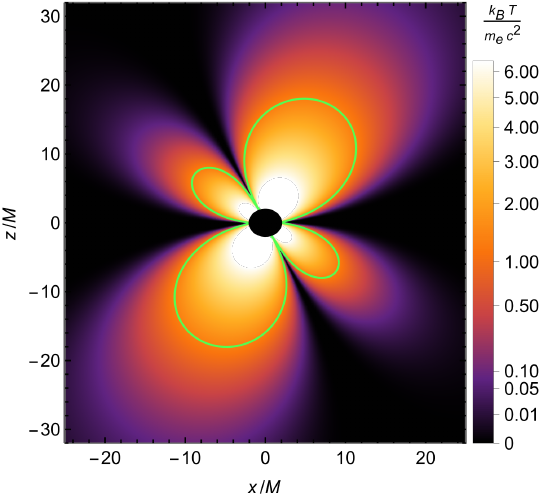}\hfill
\includegraphics[width=0.47505\textwidth]{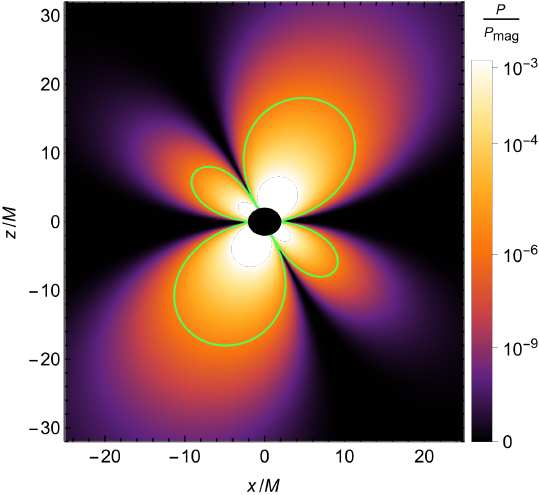}
\end{minipage}
\caption{Left panel: normalized plasma temperature $k_{B}T/(m_{e}c^{2})$. Right panel: plasma parameter $P/P_{\text{mag}}$. Both are shown in the $x$-$z$ plane of the Kerr-Schild coordinates \eqref{eq:KSx}--\eqref{eq:KSz} for $\alpha=0.5$, $\beta=400$, and $i=30^{\circ}$. The light green contour represents the dyadoregion boundary $r_{\text{d}}$ for these parameters, on which $\tilde{E}=E_{\text{c}}$.}
\label{fig:kBTandP}
\end{figure*}

\begin{figure*}[htbp]
\centering
\begin{minipage}[b]{\textwidth}
\centering
\includegraphics[width=0.475\textwidth]{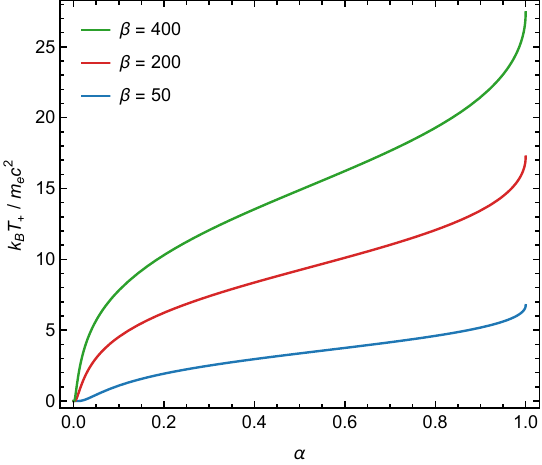}\hfill
\includegraphics[width=0.4729\textwidth]{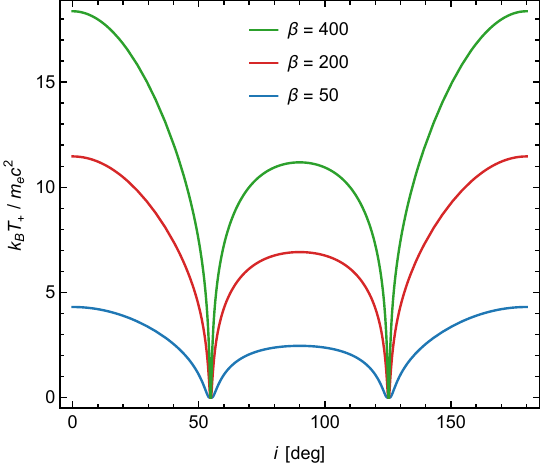}
\end{minipage}
\caption{Left: north-pole horizon temperature $k_{B}T_{+}/(m_{e}c^{2})$ as a function of the BH spin $\alpha$ for selected values of $\beta$ with fixed inclination $i=30^{\circ}$. In this example, the temperature increases monotonically with both $\alpha$ and $\beta$. Right: $k_{B}T_{+}/(m_{e}c^{2})$ versus $i$ with $\alpha=0.5$. The temperature reaches its maximum in the aligned configurations ($i=0^{\circ}$ or $180^{\circ}$). For $0<\alpha<1$, $T_{+}$ always vanishes at $i\approx 54.7^{\circ}$, $125.3^{\circ}$, $234.7^{\circ}$, and $305.3^{\circ}$. For an extremal Kerr BH ($\alpha=1$), however, two additional zeros appear at $90^\circ$ and $270^\circ$. These zeros occur where either $\tilde{E}_{+}$ or $\tilde{B}_{+}$ vanishes at the pole, i.e., when the polar axis is exactly sandwiched between two lobes of the dyadoregion, as can be inferred from Fig.~\ref{fig:dyadoregion}.}
\label{fig:kBTonEH}
\end{figure*}

For the magnetic field inclination $i=30^{\circ}$, the plasma temperature is illustrated in the left panel of Fig.~\ref{fig:kBTandP}, as an example. Clearly, the temperature distribution shows a close correspondence with the dyadoregion morphology. The high temperature inside the dyadoregion confirms that a large number of pairs are created there. Interestingly, on the dyadoregion surface (light green contour) where $\tilde{E}=E_{\text{c}}$, the temperature is almost constant,
\begin{equation}
k_{B}T_{\text{d}}\approx\frac{m_{e}c^{2}}{[6\zeta(3)]^{1/3}}\beta^{1/3}e^{-\pi/3},
\end{equation}
which is independent of the BH spin $\alpha$ and the inclination $i$. For example, $k_{B}T_{\text{d}}\approx 1.34m_{e}c^{2}$ for $\beta=400$, the field strength parameter adopted in Fig.~\ref{fig:kBTandP}. Because the temperature near the BH is not sufficiently resolved by the color map, following Cherubini et al. \cite{Cherubini:2025lnc} we plot the temperature at the north pole of the horizon in Fig.~\ref{fig:kBTonEH}, with the help of Eqs.~\eqref{eq:tildeE} and \eqref{eq:tildeB}.

\begin{figure*}[htbp]
\centering
\begin{minipage}[b]{\textwidth}
\centering
\includegraphics[width=0.475\textwidth]{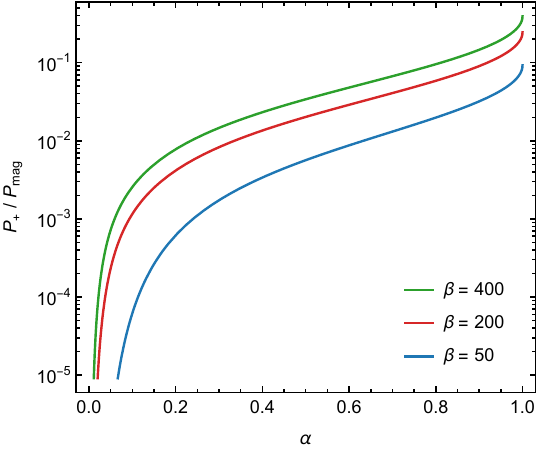}\hfill
\includegraphics[width=0.4718\textwidth]{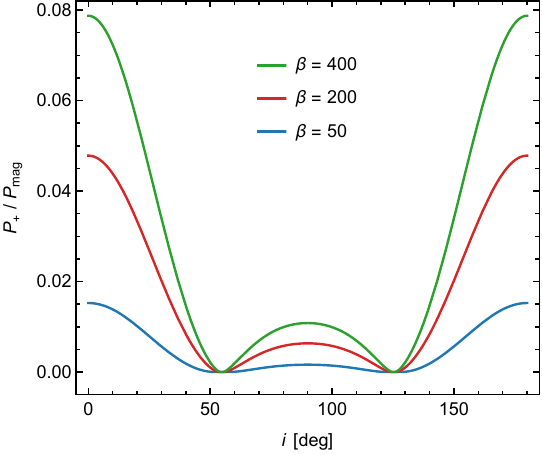}
\end{minipage}
\caption{Ratio of plasma pressure to magnetic pressure at the north pole of the horizon, $P_{+}/P_{\text{mag}}$. The parameter choice is the same as in Fig.~\ref{fig:kBTonEH}. Since $P\propto T^{4}$, the behavior of the pressure is analogous to that of the temperature.}
\label{fig:pressureonEH}
\end{figure*}

It is also instructive to compare the plasma pressure $P$ (or the energy density $\epsilon$) with the magnetic pressure $P_{\text{mag}}\approx B^{2}/(8\pi)$. As shown in Fig.~\ref{fig:kBTandP} (right panel) and Fig.~\ref{fig:pressureonEH}, for typical GRB parameters, the ratio $P/P_{\text{mag}}$ is much less than unity even at the horizon, which indicates that the magnetic field is dominant initially. Since $P/P_{\text{mag}}\propto \beta^{2/3}$, the weaker the magnetic field, the smaller the ratio, as noted in \cite{Cherubini:2025lnc}.

The initial magnetic dominance suggests that the substantial magnetic energy could be converted into plasma kinetic energy via ideal magnetohydrodynamic processes. This provides a viable mechanism for launching ultrarelativistic outflows, which are believed to be responsible for the GRB prompt emission.

%%%%%%%%%%%%%%%%%%%%%%%%%%%%%%%%%%%%%%%%%%%%%%%%%%%%%%%%%%%%%%%%%%%%%%%%%%%%
\section{Conclusions and discussions}\label{sec:conclu}
As a natural extension of the recent work by Cherubini et al. \cite{Cherubini:2025lnc}, in the present paper we systematically investigate vacuum breakdown and $e^{+}e^{-}$ pair creation around a Kerr BH immersed in an asymptotically uniform magnetic field that is inclined with respect to the BH spin axis. To this end, we employ the exact source-free Maxwell solution given by \eqref{eq:A0}--\eqref{eq:A3}, which is stationary but generally not axisymmetric in Kerr background, unlike the Wald solution \eqref{eq:Wald}.

Vacuum breakdown manifests itself only in a finite region near the BH. Based on the electromagnetic invariants \eqref{eq:scriptF} and \eqref{eq:scriptG}, we define the dyadoregion as a region outside the event horizon where the induced electric field exceeds the critical value $E_{\text{c}}=m_{e}^{2}c^{3}/(e\hbar)$ and $e^{+}e^{-}$ pairs are created. The dyadoregion surface $r_{\text{d}}$ is determined by the contour $\tilde{E}=E_{\text{c}}$, with $\tilde{E}$ being the electric field strength in a privileged frame in which the electric and magnetic fields are parallel to each other. It is evident from
Fig.~\ref{fig:dyadoregion} that the dyadoregion consists of several lobes whose number, size, and spatial orientation vary with the magnetic field inclination $i$. The observed diversity in dyadoregion morphology is a direct consequence of the loss of axisymmetry of the electromagnetic field. Notably, some lobes disappear behind the event horizon in pairs for certain ranges of $i$. We also provide an approximate analytic expression for the dyadoregion radius, Eq.~\eqref{eq:barrd}, which readers can use to reconstruct the lobe structure with sufficient accuracy for any physically allowed BH spin $\alpha$, field strength $\beta$, and inclination $i$ without having to solve the equation \eqref{eq:contour} numerically.

Fig.~\ref{fig:betamin} shows the minimum magnetic field $\beta_{\text{min}}$ required to initiate pair creation as a function of $i$ for various BH spins. For most inclinations, $\beta_{\text{min}}$ is appreciably lower than that in the aligned configuration, indicating that a misaligned magnetic field actually facilitates pair creation. Interestingly, we note that pair creation by vacuum breakdown is still possible for an extremal Kerr BH ($\alpha=1$) with a perfectly aligned field. In light of this, and given that misaligned configurations are likely the norm in realistic astrophysical settings, not aligned ones, we argue that the so-called ``BH Meissner effect''---the expulsion of stationary and axisymmetric external fields from the horizon of a maximally spinning hole---is of little relevance for the physics of vacuum breakdown. In fact, the conductive electrodynamic simulations for an extremal Kerr BH placed in an aligned uniform magnetic field \cite{Komissarov:2007rc}, as well as the more realistic GRMHD simulations incorporating magnetized accretion flows \cite{McKinney:2004ka,Komissarov:2005wj,Komissarov:2007rc,Pan:2015imp}, have consistently failed to observe the expulsion of magnetic flux. On the contrary, magnetic field lines are pulled back onto the extremal horizon due to the high conductivity of the magnetospheres.

We then estimate the electromagnetic energy $\mathcal{E}$ stored inside the dyadoregion, which provides the total energy budget for the pair creation by vacuum breakdown. It is found that the energy $\mathcal{E}$ is maximal when the magnetic field is aligned with the BH spin axis, and minimal when it is perpendicular, as shown in Fig.~\ref{fig:energyEM}. A modified scaling law for the dyadoregion energy is derived in Eq.~\eqref{eq:scalinglaw}, revealing that $\mathcal{E}\propto M^{3}\alpha^{3/2}\beta^{7/2}\eta(i)$, where $M$ is the BH mass and the geometric factor $\eta(i)$ is well approximated by a truncated cosine series with a period of $\pi$ (Eq.~\eqref{eq:etaofi}). Compared with the strong dependence on the magnetic field strength $\beta$, the variation of $\mathcal{E}$ with the inclination $i$ is relatively modest. To connect our theoretical results with GRB observations, we introduce a beaming factor $f_{b}(i)$ defined as the ratio $\Omega_{90}(i)/(4\pi)$, where $\Omega_{90}(i)$ is the minimum solid angle that contains $90\%$ of the total energy. This factor quantifies the degree to which the energy is concentrated and permits a direct conversion between the intrinsic dyadoregion energy $\mathcal{E}$ and the observed isotropic equivalent energy $\mathcal{E}_{\text{iso}}$ via Eq.~\eqref{eq:Eiso}. For stellar-mass BHs, the isotropic energies predicted by the model under study are consistent with the typical values ($10^{52}$ to $10^{54}\operatorname{erg}$) observed in GRBs \cite{Ruffini:2016jmq}.

The thermodynamic properties of the $e^{+}e^{-}\gamma$ plasma right after thermalization, including temperature, number density, energy density, and pressure, are also evaluated. From Fig.~\ref{fig:kBTandP}, a clear correspondence is observed between the distribution of thermodynamic quantities and the dyadoregion morphology, as expected. Furthermore, the plasma parameter $P/P_{\text{mag}}$ is much smaller than unity inside the dyadoregion, suggesting an initial magnetic dominance that could allow the magnetic energy reservoir to be efficiently tapped and converted into plasma kinetic energy via ideal magnetohydrodynamics. These results can serve as the initial conditions for subsequent dynamical evolution of the plasma. However, a full GRMHD simulation is far beyond the scope of the present paper. We leave this for future work.

It is worth noticing that the mass quadrupole moment of a rapidly rotating NS approaches the Kerr value as the star nears its maximum mass \cite{Cipolletta:2015nga}, so that the exterior spacetime and multipole moments tend to those of a Kerr BH. Therefore, in line with Cherubini et al. \cite{Cherubini:2025lnc}, we also expect with due caution that our main conclusions on vacuum breakdown and the dyadoregion remain qualitatively valid for such a rotating NS, provided it has a sufficiently strong magnetic field.

Throughout this work we have adopted the test field approximation, which is well justified for realistic astrophysical scenarios as argued at the beginning of Sec.~\ref{sec:testfield}. Nonetheless, from a purely theoretical perspective it would be interesting to explore vacuum breakdown in a BH spacetime where the electromagnetic field and gravity are completely coupled. A promising candidate is the Kerr-Bertotti-Robinson solution \cite{Podolsky:2025tle}, an exact Einstein-Maxwell spacetime that describes a Kerr BH immersed in an external uniform magnetic (or electric) field aligned with its rotation axis. In that case, the magnetic field parameter enters the metric functions, and one may expect that the BH mass will affect the dyadoregion morphology in a nontrivial way.
%%%%%%%%%%%%%%%%%%%%%%%%%%%%%%%%%%%%%%%%%%%%%%%%%%%%%%%%%%%%%%%%%%%%%%%%%%%%
\begin{acknowledgments}
The first author, Ruixin Yang, would like to thank Jorge A. Rueda for his generous and helpful correspondence. This work was supported by the National Natural Science Foundation of China under Grant No.~12275078, 11875026, 12035005, 2020YFC2201400, and the innovative research group of Hunan Province under Grant No.~2024JJ1006.
\end{acknowledgments}
%%%%%%%%%%%%%%%%%%%%%%%%%%%%%%%%%%%%%%%%%%%%%%%%%%%%%%%%%%%%%%%%%%%%%%%%%%%%
%\appendix
%\section{Appendix}
%\clearpage
%%%%%%%%%%%%%%%%%%%%%%%%%%%%%%%%%%%%%%%%%%%%%%%%%%%%%%%%%%%%%%%%%%%%%%%%%%%%
\bibliography{mmKref}

@article{Schwinger:1951nm,
    author = "Schwinger, Julian S.",
    editor = "Milton, K. A.",
    title = "{On gauge invariance and vacuum polarization}",
    doi = "10.1103/PhysRev.82.664",
    journal = "Phys. Rev.",
    volume = "82",
    pages = "664--679",
    year = "1951"
}

@article{Ruffini:2009hg,
    author = "Ruffini, Remo and Vereshchagin, Gregory and Xue, She-Sheng",
    title = "{Electron-positron pairs in physics and astrophysics: from heavy nuclei to black holes}",
    doi = "10.1016/j.physrep.2009.10.004",
    journal = "Phys. Rept.",
    volume = "487",
    pages = "1--140",
    year = "2010"
}

@article{Popham:1998ab,
    author = "Popham, Robert and Woosley, S. E. and Fryer, Chris",
    title = "{Hyperaccreting black holes and gamma-ray bursts}",
    doi = "10.1086/307259",
    journal = "Astrophys. J.",
    volume = "518",
    pages = "356--374",
    year = "1999"
}

@article{Narayan:2001qi,
    author = "Narayan, Ramesh and Piran, Tsvi and Kumar, Pawan",
    title = "{Accretion models of gamma-ray bursts}",
    doi = "10.1086/322267",
    journal = "Astrophys. J.",
    volume = "557",
    pages = "949",
    year = "2001"
}

@article{Kohri:2002kz,
    author = "Kohri, Kazunori and Mineshige, Shin",
    title = "{Can neutrino cooled accretion disk be an origin of gamma-ray bursts?}",
    doi = "10.1086/342166",
    journal = "Astrophys. J.",
    volume = "577",
    pages = "311--321",
    year = "2002"
}

@article{DiMatteo:2002iex,
    author = "Di Matteo, Tiziana and Perna, Rosalba and Narayan, Ramesh",
    title = "{Neutrino trapping and accretion models for gamma-ray bursts}",
    doi = "10.1086/342832",
    journal = "Astrophys. J.",
    volume = "579",
    pages = "706--715",
    year = "2002"
}

@article{Kohri:2005tq,
    author = "Kohri, Kazunori and Narayan, Ramesh and Piran, Tsvi",
    title = "{Neutrino-dominated accretion and supernovae}",
    doi = "10.1086/431354",
    journal = "Astrophys. J.",
    volume = "629",
    pages = "341--361",
    year = "2005"
}

@article{Lee:2005se,
    author = "Lee, William H. and Ramirez-Ruiz, Enrico and Page, Dany",
    title = "{Dynamical evolution of neutrino-cooled accretion disks: Detailed microphysics, lepton-driven convection, and global energetics}",
    doi = "10.1086/432373",
    journal = "Astrophys. J.",
    volume = "632",
    pages = "421--437",
    year = "2005"
}

@article{Gu:2006nu,
    author = "Gu, Wei-Min and Liu, Tong and Lu, Ju-Fu",
    title = "{Neutrino-Dominated Accretion Models for Gamma-Ray Bursts: Effects of General Relativity and Neutrino Opacity}",
    doi = "10.1086/505140",
    journal = "Astrophys. J. Lett.",
    volume = "643",
    pages = "L87--L90",
    year = "2006"
}

@article{Chen:2006rra,
    author = "Chen, Wen-Xin and Beloborodov, Andrei M.",
    title = "{Neutrino-Cooled Accretion Disks around Spinning Black Hole}",
    doi = "10.1086/508923",
    journal = "Astrophys. J.",
    volume = "657",
    pages = "383--399",
    year = "2007"
}

@article{Kawanaka:2007sb,
    author = "Kawanaka, Norita and Mineshige, Shin",
    title = "{Neutrino Cooled Disk and Its Stability}",
    doi = "10.1086/517985",
    journal = "Astrophys. J.",
    volume = "662",
    pages = "1156--1166",
    year = "2007"
}

@article{Janiuk:2009gc,
    author = "Janiuk, Agnieszka and Yuan, Ye-Fei",
    title = "{The role of black hole spin and magnetic field threading the unstable neutrino disk in Gamma Ray Bursts}",
    doi = "10.1051/0004-6361/200912725",
    journal = "Astron. Astrophys.",
    volume = "509",
    pages = "A55",
    year = "2010"
}

@article{Kawanaka:2012ub,
    author = "Kawanaka, Norita and Piran, Tsvi and Krolik, Julian H.",
    title = "{Jet Luminosity From Neutrino-Dominated Accretion Flows in Gamma-Ray Bursts}",
    doi = "10.1088/0004-637X/766/1/31",
    journal = "Astrophys. J.",
    volume = "766",
    pages = "31",
    year = "2013"
}

@article{Luo:2013zx,
    author = "Luo, Shu and Yuan, Feng",
    title = "{Global Neutrino Heating in Hyperaccretion Flows}",
    doi = "10.1093/mnras/stt337",
    journal = "Mon. Not. Roy. Astron. Soc.",
    volume = "431",
    pages = "2362",
    year = "2013"
}

@article{Xue:2013boa,
    author = "Xue, Li and Liu, Tong and Gu, Wei-Min and Lu, Ju-Fu",
    title = "{Relativistic global solutions of neutrino-dominated accretion flows}",
    doi = "10.1088/0067-0049/207/2/23",
    journal = "Astrophys. J. Suppl.",
    volume = "207",
    pages = "23",
    year = "2013"
}

@article{Liu:2017kga,
    author = "Liu, Tong and Gu, Wei-Min and Zhang, Bing",
    title = "{Neutrino-dominated accretion flows as the central engine of gamma-ray bursts}",
    doi = "10.1016/j.newar.2017.07.001",
    journal = "New Astron. Rev.",
    volume = "79",
    pages = "1--25",
    year = "2017"
}

@article{Uribe:2019cpq,
    author = "Uribe, J. D. and Becerra-Vergara, Eduar Antonio and Rueda, Jorge Armando",
    title = "{Neutrino oscillations in a neutrino-dominated accretion disk around a Kerr BH}",
    doi = "10.3390/universe7010007",
    journal = "Universe",
    volume = "7",
    number = "1",
    pages = "7",
    year = "2021"
}

@article{Aksenov:0227,
  title = {Thermalization of Nonequilibrium Electron-Positron-Photon Plasmas},
  author = {Aksenov, A. G. and Ruffini, R. and Vereshchagin, G. V.},
  journal = {Phys. Rev. Lett.},
  volume = {99},
  issue = {12},
  pages = {125003},
  numpages = {4},
  year = {2007},
  doi = {10.1103/PhysRevLett.99.125003}
}

@article{Aksenov:2009dy,
    author = "Aksenov, A. G. and Ruffini, R. and Vereshchagin, G. V.",
    title = "{Thermalization of the mildly relativistic plasma}",
    doi = "10.1103/PhysRevD.79.043008",
    journal = "Phys. Rev. D",
    volume = "79",
    pages = "043008",
    year = "2009"
}

@article{Preparata:1998rz,
    author = "Preparata, Giuliano and Ruffini, Remo and Xue, She-Sheng",
    title = "{The Dyadosphere of black holes and gamma-ray bursts}",
    eprint = "astro-ph/9810182",
    archivePrefix = "arXiv",
    journal = "Astron. Astrophys.",
    volume = "338",
    pages = "L87--L90",
    year = "1998"
}

@article{Cherubini:2009ww,
    author = "Cherubini, C. and Geralico, A. and Rueda H., J. A. and Ruffini, R.",
    title = "{$e^-e^+$ pair creation by vacuum polarization around electromagnetic black holes}",
    doi = "10.1103/PhysRevD.79.124002",
    journal = "Phys. Rev. D",
    volume = "79",
    pages = "124002",
    year = "2009"
}

@article{Wald:1974np,
    author = "Wald, Robert M.",
    title = "{Black hole in a uniform magnetic field}",
    doi = "10.1103/PhysRevD.10.1680",
    journal = "Phys. Rev. D",
    volume = "10",
    pages = "1680--1685",
    year = "1974"
}

@article{Moradi:2021hus,
    author = "Moradi, R. and Rueda, J. A. and Ruffini, R. and Li, Liang and Bianco, C. L. and Campion, S. and Cherubini, C. and Filippi, S. and Wang, Y. and Xue, S. S.",
    title = "{Nature of the ultrarelativistic prompt emission phase of GRB 190114C}",
    doi = "10.1103/PhysRevD.104.063043",
    journal = "Phys. Rev. D",
    volume = "104",
    number = "6",
    pages = "063043",
    year = "2021"
}

@article{Rastegarnia:2022rds,
    author = "Rastegarnia, F. and Moradi, R. and Rueda, J. A. and Ruffini, R. and Li, Liang and Eslamzadeh, S. and Wang, Y. and Xue, S. S.",
    title = "{The structure of the ultrarelativistic prompt emission phase and the properties of the black hole in GRB 180720B}",
    doi = "10.1140/epjc/s10052-022-10750-x",
    journal = "Eur. Phys. J. C",
    volume = "82",
    number = "9",
    pages = "778",
    year = "2022"
}

@article{Cherubini:2025lnc,
    author = "Cherubini, C. and Moradi, R. and Rueda, J. A. and Ruffini, R.",
    title = "{Vacuum breakdown around a Kerr black hole surrounded by a magnetic field}",
    doi = "10.1051/0004-6361/202556413",
    journal = "Astron. Astrophys.",
    volume = "704",
    pages = "A69",
    year = "2025"
}

@article{Obergaulinger:2008pb,
    author = "Obergaulinger, M. and Cerda-Duran, P. and Muller, E. and Aloy, M. A.",
    title = "{Semi-global simulations of the magneto-rotational instability in core collapse supernovae}",
    doi = "10.1051/0004-6361/200811323",
    journal = "Astron. Astrophys.",
    volume = "498",
    pages = "241",
    year = "2009"
}

@article{Kiuchi:2014hja,
    author = "Kiuchi, Kenta and Kyutoku, Koutarou and Sekiguchi, Yuichiro and Shibata, Masaru and Wada, Tomohide",
    title = "{High resolution numerical-relativity simulations for the merger of binary magnetized neutron stars}",
    doi = "10.1103/PhysRevD.90.041502",
    journal = "Phys. Rev. D",
    volume = "90",
    pages = "041502",
    year = "2014"
}

@article{Kiuchi:2015sga,
    author = "Kiuchi, Kenta and Cerd{\'a}-Dur{\'a}n, Pablo and Kyutoku, Koutarou and Sekiguchi, Yuichiro and Shibata, Masaru",
    title = "{Efficient magnetic-field amplification due to the Kelvin-Helmholtz instability in binary neutron star mergers}",
    doi = "10.1103/PhysRevD.92.124034",
    journal = "Phys. Rev. D",
    volume = "92",
    number = "12",
    pages = "124034",
    year = "2015"
}

@article{Obergaulinger:2017qno,
    author = "Obergaulinger, Martin and Aloy, Miguel {\'A}ngel",
    title = "{Protomagnetar and black hole formation in high-mass stars}",
    doi = "10.1093/mnrasl/slx046",
    journal = "Mon. Not. Roy. Astron. Soc.",
    volume = "469",
    number = "1",
    pages = "L43--L47",
    year = "2017"
}

@article{Proga:2003ap,
    author = "Proga, D. and Begelman, M. C.",
    title = "{Accretion of low angular momentum material onto black holes: 2-D magnetohydrodynamical case}",
    doi = "10.1086/375773",
    journal = "Astrophys. J.",
    volume = "592",
    pages = "767--781",
    year = "2003"
}

@article{Liska:2017alm,
    author = "Liska, Matthew and Hesp, Casper and Tchekhovskoy, Alexander and Ingram, Adam and van der Klis, Michiel and Markoff, Sera",
    title = "{Formation of Precessing Jets by Tilted Black-hole Discs in 3D General Relativistic MHD Simulations}",
    doi = "10.1093/mnrasl/slx174",
    journal = "Mon. Not. Roy. Astron. Soc.",
    volume = "474",
    number = "1",
    pages = "L81--L85",
    year = "2018"
}

@article{McKinney:2012wd,
    author = "McKinney, Jonathan C. and Tchekhovskoy, Alexander and Blandford, Roger D.",
    title = "{Alignment of Magnetized Accretion Disks and Relativistic Jets with Spinning Black Holes}",
    doi = "10.1126/science.1230811",
    journal = "Science",
    volume = "339",
    pages = "49--52",
    year = "2013"
}

@article{El-Badry:2023pah,
    author = "El-Badry, Kareem and others",
    title = "{A red giant orbiting a black hole}",
    doi = "10.1093/mnras/stad799",
    journal = "Mon. Not. Roy. Astron. Soc.",
    volume = "521",
    number = "3",
    pages = "4323--4348",
    year = "2023"
}

@article{Barkov:2012sj,
    author = "Barkov, M. V. and Khangulyan, D. V. and Popov, S. B.",
    title = "{Jets and gamma-ray emission from isolated accreting black holes}",
    doi = "10.1111/j.1365-2966.2012.22029.x",
    journal = "Mon. Not. Roy. Astron. Soc.",
    volume = "427",
    pages = "589--594",
    year = "2012"
}

@article{Kin:2025axi,
    author = "Kin, Koki and Kuze, Riku and Kimura, Shigeo S.",
    title = "{Galactic Isolated Stellar-mass Black Holes with the Magnetospheric Spark Gap as Possible GeV{\textendash}TeV Gamma-Ray Unidentified Sources}",
    doi = "10.3847/1538-4357/adcb3d",
    journal = "Astrophys. J.",
    volume = "985",
    number = "2",
    pages = "251",
    year = "2025"
}

@article{Figueiredo:2025xbo,
    author = "Figueiredo, Enzo and Cerutti, Beno{\^\i}t and Parfrey, Kyle",
    title = "{Effect of magnetic field inclination on black hole jet power and particle acceleration}",
    doi = "10.1051/0004-6361/202555826",
    journal = "Astron. Astrophys.",
    volume = "700",
    pages = "L19",
    year = "2025"
}

@Article{Bicak1976,
author={Bi{\v{c}}{\'a}k, J.
and Dvo{\v{r}}{\'a}k, L.},
title={Stationary electromagnetic fields around black holes. II. General solutions and the fields of some special sources near a Kerr black hole},
journal={Gen. Rel. Grav.},
year={1976},
month={Dec},
day={01},
volume={7},
number={12},
pages={959-983},
doi={10.1007/BF00766421}
}

@article{Bicak1985,
    author = {Bi{\v{c}}{\'a}k, Jiří and Jani{\v{s}}, Václav},
    title = {Magnetic fluxes across black holes},
    journal = {Mon. Not. Roy. Astron. Soc.},
    volume = {212},
    number = {4},
    pages = {899-915},
    year = {1985},
    month = {02},
    doi = {10.1093/mnras/212.4.899}
}

@article{Karas:2008xj,
    author = "Karas, V. and Kopacek, O.",
    title = "{Magnetic layers and neutral points near rotating black hole}",
    doi = "10.1088/0264-9381/26/2/025004",
    journal = "Class. Quant. Grav.",
    volume = "26",
    pages = "025004",
    year = "2009"
}

@article{Karas:2012mp,
    author = "Karas, V. and Kopacek, O. and Kunneriath, D.",
    title = "{Influence of frame-dragging on magnetic null points near rotating black hole}",
    doi = "10.1088/0264-9381/29/3/035010",
    journal = "Class. Quant. Grav.",
    volume = "29",
    pages = "035010",
    year = "2012"
}

@article{Kopacek:2014moa,
    author = "Kop{\'a}{\v{c}}ek, O. and Karas, V.",
    title = "{Inducing chaos by breaking axial symmetry in a black hole magnetosphere}",
    doi = "10.1088/0004-637X/787/2/117",
    journal = "Astrophys. J.",
    volume = "787",
    pages = "117",
    year = "2014"
}

@article{Kopacek:2014ooa,
    author = "Kop{\'a}{\v{c}}ek, Ond{\v{r}}ej and Karas, Vladim{\'\i}r",
    title = "{Regular and chaotic motion in general relativity: The case of an inclined black hole magnetosphere}",
    doi = "10.1088/1742-6596/600/1/012070",
    journal = "J. Phys. Conf. Ser.",
    volume = "600",
    number = "1",
    pages = "012070",
    year = "2015"
}

@article{Bicak:2023rsz,
    author = "Bi{\v{c}}{\'a}k, Ji{\v{r}}{\'\i} and Kubiz{\v{n}}{\'a}k, David and Perche, T. Rick",
    title = "{Migrating Carrollian particles on magnetized black hole horizons}",
    doi = "10.1103/PhysRevD.107.104014",
    journal = "Phys. Rev. D",
    volume = "107",
    number = "10",
    pages = "104014",
    year = "2023"
}

@article{Karas:1989zz,
    author = "Karas, Vladimir",
    title = "{Asymptotically uniform magnetic field near a Kerr black hole}",
    doi = "10.1103/PhysRevD.40.2121",
    journal = "Phys. Rev. D",
    volume = "40",
    pages = "2121--2123",
    year = "1989"
}

@article{Punsly:1989zz,
    author = "Punsly, Brian and Coroniti, Ferdinand V.",
    title = "{Electrodynamics of the event horizon}",
    doi = "10.1103/PhysRevD.40.3834",
    journal = "Phys. Rev. D",
    volume = "40",
    pages = "3834--3857",
    year = "1989"
}

@article{Dovciak_2000,
doi = {10.1088/0143-0807/21/4/304},
year = {2000},
month = {jul},
publisher = {},
volume = {21},
number = {4},
pages = {303},
author = {M Dovciak and V Karas and A Lanza},
title = {Magnetic fields around black holes},
journal = {Eur. J. Phys.}
}

@article{Neronov:2007vy,
    author = "Neronov, A. and Aharonian, Felix A.",
    title = "{Production of TeV gamma-radiation in the vicinity of the supermassive black hole in the giant radiogalaxy M87}",
    doi = "10.1086/522199",
    journal = "Astrophys. J.",
    volume = "671",
    pages = "85",
    year = "2007"
}

@article{Neronov:2009zz,
    author = "Neronov, A. Yu and Semikoz, D. V. and Tkachev, I. I.",
    title = "{Ultra-High Energy Cosmic Ray production in the polar cap regions of black hole magnetospheres}",
    doi = "10.1088/1367-2630/11/6/065015",
    journal = "New J. Phys.",
    volume = "11",
    pages = "065015",
    year = "2009"
}

@article{Kalashev:2012cm,
    author = "Kalashev, O. E. and Ptitsyna, K. V. and Troitsky, S. V.",
    title = "{Towards a model of population of astrophysical sources of ultra-high-energy cosmic rays}",
    doi = "10.1103/PhysRevD.86.063005",
    journal = "Phys. Rev. D",
    volume = "86",
    pages = "063005",
    year = "2012"
}

@article{Tursunov:2019mox,
    author = "Tursunov, Arman and Zaja{\v{c}}ek, Michal and Eckart, Andreas and Kolo{\v{s}}, Martin and Britzen, Silke and Stuchl{\'\i}k, Zden{\v{e}}k and Czerny, Bozena and Karas, Vladim{\'\i}r",
    title = "{Effect of Electromagnetic Interaction on Galactic Center Flare Components}",
    doi = "10.3847/1538-4357/ab980e",
    journal = "Astrophys. J.",
    volume = "897",
    number = "1",
    pages = "99",
    year = "2020"
}

@article{Karas:2020ixz,
    author = "Karas, Vladimir and Kopacek, Ondrej",
    title = "{Near-horizon structure of escape zones of electrically charged particles around weakly magnetized rotating black hole: Case of oblique magnetosphere}",
    doi = "10.1002/asna.202113934",
    journal = "Astron. Nachr.",
    volume = "342",
    number = "1-2",
    pages = "357--363",
    year = "2021"
}

@article{Kopacek:2020scv,
    author = "Kop{\'a}{\v{c}}ek, Ond{\v{r}}ej and Karas, Vladim{\'\i}r",
    title = "{Near-horizon structure of escape zones of electrically charged particles around weakly magnetized rotating black hole. II. Acceleration and escape in the oblique magnetosphere}",
    doi = "10.3847/1538-4357/ababa8",
    journal = "Astrophys. J.",
    volume = "900",
    number = "2",
    pages = "119",
    year = "2020"
}

@article{Ressler:2021jjr,
    author = "Ressler, Sean M. and Quataert, Eliot and White, Christopher J. and Blaes, Omer",
    title = "{Magnetically modified spherical accretion in GRMHD: reconnection-driven convection and jet propagation}",
    doi = "10.1093/mnras/stab311",
    journal = "Mon. Not. Roy. Astron. Soc.",
    volume = "504",
    number = "4",
    pages = "6076--6095",
    year = "2021"
}

@article{Chen:2021sya,
    author = "Chen, K. and Dai, Z. G.",
    title = "{Charging and Electromagnetic Radiation during the Inspiral of a Black Hole{\textendash}Neutron Star Binary}",
    doi = "10.3847/1538-4357/abd7a7",
    journal = "Astrophys. J.",
    volume = "909",
    number = "1",
    pages = "4",
    year = "2021"
}

@article{Hu:2021paa,
    author = "Hu, Shiyang and Wu, Xin and Liang, Enwei",
    title = "{Construction of a Second-order Six-dimensional Hamiltonian-conserving Scheme}",
    doi = "10.3847/1538-4365/ac1ff3",
    journal = "Astrophys. J. Supp.",
    volume = "257",
    number = "2",
    pages = "40",
    year = "2021"
}

@article{James:2024bib,
    author = "James, Bestin and Janiuk, Agnieszka and Karas, Vladimir",
    title = "{Black hole outflows initiated by a large-scale magnetic field - Effects of aligned and misaligned spin}",
    doi = "10.1051/0004-6361/202349134",
    journal = "Astron. Astrophys.",
    volume = "687",
    pages = "A185",
    year = "2024"
}

@article{Frolov:2024xyo,
    author = "Frolov, Valeri P.",
    title = "{Motion of a rotating black hole in a homogeneous electromagnetic field}",
    doi = "10.1103/PhysRevD.109.064045",
    journal = "Phys. Rev. D",
    volume = "109",
    number = "6",
    pages = "064045",
    year = "2024"
}

@article{Frolov:2024tiu,
    author = "Frolov, Valeri P. and Koek, Alex",
    title = "{Motion of a weakly charged rotating black hole in a homogeneous electromagnetic field}",
    doi = "10.1103/PhysRevD.110.064052",
    journal = "Phys. Rev. D",
    volume = "110",
    number = "6",
    pages = "064052",
    year = "2024"
}

@article{Tursunov:2019oiq,
    author = "Tursunov, Arman and Dadhich, Naresh",
    title = "{Fifty years of energy extraction from rotating black hole: revisiting magnetic Penrose process}",
    doi = "10.3390/universe5050125",
    journal = "Universe",
    volume = "5",
    number = "5",
    pages = "125",
    year = "2019"
}

@article{Galtsov:1978ag,
    author = "Galtsov, D. V. and Petukhov, V. I.",
    title = "{Black Hole in an External Magnetic Field}",
    journal = "Zh. Eksp. Teor. Fiz.",
    volume = "74",
    pages = "801--818",
    year = "1978"
}

@incollection{LANDAU197543,
title = {CHAPTER 3 - CHARGES IN ELECTROMAGNETIC FIELDS},
booktitle = {The Classical Theory of Fields (Fourth Edition)},
publisher = {Pergamon},
edition = {Fourth Edition},
address = {Amsterdam},
pages = {43-65},
year = {1975},
volume = {2},
series = {Course of Theoretical Physics},
doi = {10.1016/B978-0-08-025072-4.50010-1},
author = {L. D. Landau and E. M. Lifshitz}
}

@article{Carter:1968ks,
    author = "Carter, B.",
    title = "{Hamilton-Jacobi and Schrodinger separable solutions of Einstein's equations}",
    doi = "10.1007/BF03399503",
    journal = "Commun. Math. Phys.",
    volume = "10",
    number = "4",
    pages = "280--310",
    year = "1968"
}

@article{Damour:1974qv,
    author = "Damour, Thibault and Ruffini, Remo",
    title = "{Quantum Electrodynamical Effects in Kerr-Newman Geometries}",
    reportNumber = "Print-75-0106 (IAS,PRINCETON)",
    doi = "10.1103/PhysRevLett.35.463",
    journal = "Phys. Rev. Lett.",
    volume = "35",
    pages = "463",
    year = "1975"
}

@article{McKinney:2004ka,
    author = "McKinney, Jonathan C. and Gammie, Charles F.",
    title = "{A Measurement of the electromagnetic luminosity of a Kerr black hole}",
    doi = "10.1086/422244",
    journal = "Astrophys. J.",
    volume = "611",
    pages = "977--995",
    year = "2004"
}

@article{Komissarov:2005wj,
    author = "Komissarov, S. S.",
    title = "{Observations of the Blandford-Znajek and the MHD Penrose processes in computer simulations of black hole magnetospheres}",
    doi = "10.1111/j.1365-2966.2005.08974.x",
    journal = "Mon. Not. Roy. Astron. Soc.",
    volume = "359",
    pages = "801--808",
    year = "2005"
}

@article{Komissarov:2007rc,
    author = "Komissarov, S. S. and McKinney, J. C.",
    title = "{Meissner effect and Blandford-Znajek mechanism in conductive black hole magnetospheres}",
    doi = "10.1111/j.1745-3933.2007.00301.x",
    journal = "Mon. Not. Roy. Astron. Soc.",
    volume = "377",
    pages = "L49--L53",
    year = "2007"
}

@article{Pan:2015imp,
    author = "Pan, Zhen and Yu, Cong",
    title = "{Analytic Properties of Force-free Jets in the Kerr Spacetime{\textemdash}II}",
    doi = "10.3847/0004-637X/816/2/77",
    journal = "Astrophys. J.",
    volume = "816",
    number = "2",
    pages = "77",
    year = "2016"
}

@article{Ruffini:2016jmq,
    author = "Ruffini, R. and others",
    title = "{On the classification of GRBs and their occurrence rates}",
    doi = "10.3847/0004-637X/832/2/136",
    journal = "Astrophys. J.",
    volume = "832",
    number = "2",
    pages = "136",
    year = "2016"
}

@article{Cipolletta:2015nga,
    author = "Cipolletta, Federico and Cherubini, Christian and Filippi, Simonetta and Rueda, Jorge Armando and Ruffini, Remo",
    title = "{Fast Rotating Neutron Stars with Realistic Nuclear Matter Equation of State}",
    doi = "10.1103/PhysRevD.92.023007",
    journal = "Phys. Rev. D",
    volume = "92",
    number = "2",
    pages = "023007",
    year = "2015"
}

@article{Podolsky:2025tle,
    author = "Podolsky, Jiri and Ovcharenko, Hryhorii",
    title = "{Kerr Black Hole in a Uniform Bertotti-Robinson Magnetic Field: An Exact Solution}",
    doi = "10.1103/rfgv-ybz5",
    journal = "Phys. Rev. Lett.",
    volume = "135",
    number = "18",
    pages = "181401",
    year = "2025"
}
%\bibliographystyle{apsrev4-2}
%%%%%%%%%%%%%%%%%%%%%%%%%%%%%%%%%%%%%%%%%%%%%%%%%%%%%%%%%%%%%%%%%%%%%%%%%%%%
\end{document}